\documentclass[11pt]{iopart}

\usepackage{graphicx}

\expandafter\let\csname equation*\endcsname\relax

\expandafter\let\csname endequation*\endcsname\relax

\usepackage{mathtools}
\usepackage{float}
\usepackage{microtype}
\usepackage{tabularx}

\begin{document}



\title[The role of plasma-molecule interactions on detachment]{The role of plasma-molecule interactions on power and particle balance during detachment on the TCV tokamak}



\author{K. Verhaegh$^{1,2,3}$, B. Lipschultz$^2$, J.R. Harrison$^1$, B.P. Duval$^3$, A. Fil$^{1}$, M. Wensing$^3$, C. Bowman$^2$, D.S. Gahle$^{4,1}$, A. Kukushkin$^{5,6}$, D. Moulton$^1$, A. Perek$^7$, A. Pshenov$^{5,6}$, F. Federici$^2$, O. F\'{e}vrier$^3$, O. Myatra$^2$,  A. Smolders$^3$, C. Theiler$^3$, the TCV Team$^{*}$ and the EuroFusion MST1 team$^{**}$}
\address{$^1$ Culham Centre for Fusion Energy, Culham, United Kingdom} 
\address{$^2$ York Plasma Institute, University of York, United Kingdom}
\address{$^3$ Swiss Plasma Centre, \'{E}cole Polytechnique F\'{e}d\'{e}rale de Lausanne, Lausanne, Switzerland}
\address{$^4$ SUPA, University of Strathclyde, Glasgow, United Kingdom}
\address{$^5$ NCR Kurchatov Institute, Moscow, Russian Federation}
\address{$^6$ National Research Nuclear University MEphI, Moscow, Russian Federation}
\address{$^7$ DIFFER, Eindhoven, The Netherlands}
\address{$^*$ See author list of "S. Coda et al 2019 Nucl. Fusion 59 112023"}
\address{$^**$ See author list of "B. Labit et al 2019 Nucl. Fusion 59 086020"}


\ead{kevin.verhaegh@ukaea.uk}

\begin{abstract}
This paper shows experimental results from the TCV tokamak that indicate plasma-molecule interactions involving $D_2^+$ and possibly $D^-$ play an important role as sinks of energy (through hydrogenic radiation as well as dissociation) and particles during divertor detachment if low target temperatures ($< 3$ eV) are achieved. Both molecular activated recombination (MAR) and ion source reduction due to a power limitation effect are shown to be important in reducing the ion target flux during a density ramp. In contrast, the electron-ion recombination (EIR) ion sink is too small to play an important role in reducing the ion target flux. MAR or EIR do not occur during $N_2$ seeding induced detachment as the target temperatures are not sufficiently low. 

The impact of $D_2^+$ is shown to be underestimated in present (vibrationally unresolved) SOLPS-ITER simulations, which could result from an underestimated $D_2 + D^+ \rightarrow D_2^+ + D$ rate. The converged SOLPS-ITER simulations are post-processed with alternative reaction rates, resulting in considerable contributions of $D_2^+$ to particle and power losses as well as dissociation below the $D_2$ dissociation area. Those findings are in quantitative agreement with the experimental results. 
\end{abstract}


\section{Introduction}
\label{ch:introduction}

Power exhaust is a major challenge for future fusion devices \cite{Loarte2007,Wenninger2014}. The material limits for the (plasma) heat flux to the target are in the order of 5-20 $\textrm{MW/m}^2$, depending on how often the divertor should be replaced and what the expected non-plasma (e.g. neutrals and radiation) heat load is, which would be exceeded by an order of magnitude in a fusion reactor without mitigation \cite{Loarte2007,Wenninger2014,Pitts2013,Pitts2019,Escourbiac2019}. Divertor heat flux loading ($q_t$) as well as target temperature ($T_t$) can be tempered by increasing the density and/or inducing radiative losses through seeding extrinsic impurities, which will be required on reactors \cite{Loarte2007,Wenninger2014,Pitts2013}. Equation \ref{eq:qt} uses the plasma sheath equation \cite{Stangeby2000} and indicates that if target pressure ($p_t$) losses do not occur, reducing the target temperature results in: 1) a decrease of the kinetic part of the target heat flux loading ($q_t^{kin} \propto p_t \gamma T_t^{1/2}$ where $\gamma \approx 7$ is the sheath heat transfer coefficient); 2) an increase in the potential surface recombination target heat flux loading ($q_t^{pot} \propto p_t \epsilon T_t^{-1/2}$ where $\epsilon = 13.6$ eV is the ionisation potential). Minimising equation \ref{eq:qt} with respect to $T_t$ and comparing to a typical sheath-limited $T_t$ of 100 eV, we see that the total (plasma) target heat flux load may only be reduced by a factor 4-5 without target pressure losses.  

\begin{equation}
\begin{split}
    q_t &= \underbrace{\Gamma_t \gamma T_t}_{q_t^{kin}} + \underbrace{\Gamma_t \epsilon}_{q_t^{pot}} \\
   \Gamma_t &\propto p_t / T_t^{1/2} \\
   \rightarrow &q_t \propto p_t (\gamma T_t^{1/2} + \epsilon T_t^{-1/2})
\label{eq:qt}
\end{split}
\end{equation}

\subsection{Detachment physics: power, particle and momentum balance}

Divertor detachment can facilitate larger heat flux reductions as it results in simultaneous: 1) power losses, 2) target pressure losses (through volumetric momentum losses \cite{Loarte2007,Pitcher1997,Stangeby2000}, which are preferable for maintaining a sufficient core performance in a reactor, and upstream pressure losses \cite{Verhaegh2019}); 3) target particle flux losses either through a reduction of the divertor ionisation source and/or an increase in the volumetric ion recombination sink \cite{Krasheninnikov1997,Lipschultz1999}. These losses are brought on by many plasma-atom and molecular interactions at relatively low target temperatures ($T_t < 5$ eV) \cite{Stangeby2000}. Below we briefly highlight the key aspects of detachment in terms of power, particle and momentum balance.

Divertor particle balance is summarised in equation \ref{eq:PartBal}, where the ion flux reaching the target equals the ion source in the divertor $\Gamma_i$ minus the recombination ion sinks in the divertor $\Gamma_r$ plus a 'net' inflow of ions (positive - implies a net flow towards the target, negative - the converse) $\Gamma_u$. 

\begin{equation}
    \Gamma_t = \Gamma_i - \Gamma_r + \Gamma_u
    \label{eq:PartBal}
\end{equation}

In the 'high-recycling regime' (e.g. $\Gamma_t \gg \Gamma_u$), the  divertor is, by definition , an almost 'self-contained' region where the ion flux reaching the target is mainly \emph{generated} in the divertor through ionisation \cite{Lipschultz1999,Krasheninnikov2017,Pshenov2017}. Particle balance is strongly co-dependent with power balance in this regime as ionisation has an energy cost $E_{ion}$ per ionisation event. This includes radiative energy losses from excitation collisions preceding ionisation ($E_{ion}^{rad}$) together with the potential energy required to ionise a neutral into an ion ($\epsilon$). The heat flux entering the divertor $q_{div}$ is reduced by impurity radiation $q_{rad}^{imp}$. A part of the remainder of the heat flux ($q_{recl}$) is used for ionisation ($E_{ion} \Gamma_i$). After ionisation, the remainder of the heat flux is deposited at the target as particles with their kinetic energy ($\Gamma_t \gamma T_t$) - equation \ref{eq:PowerBal}. 

\begin{align}
        q_{recl} &\equiv q_{div} - q_{rad}^{imp} \\
        q_{recl} &= E_{ion} \Gamma_i + \Gamma_t \gamma T_t
        \label{eq:PowerBal}
\end{align}

Combining equations \ref{eq:PartBal} and \ref{eq:PowerBal} provides a model for the ion target flux - equation \ref{eq:PowerPartBal}, where $\Gamma_t \gg \Gamma_u$ is assumed. 

\begin{equation}
    \Gamma_t = (\frac{q_{recl}}{E_{ion}} - \Gamma_r) \times \frac{1}{1 + \frac{\gamma T_t}{E_{ion}}} 
    \label{eq:PowerPartBal}
\end{equation}


As the target temperature becomes small $\frac{\gamma T_t}{E_{ion}} \ll 1$, $\Gamma_t \approx \frac{q_{recl}}{E_{ion}} - \Gamma_r$, so that the ion target flux can be reduced through: 1) a reduction of $q_{recl}$; 2) an increase of the energy cost of ionisation $E_{ion}$; 3) volumetric recombination $\Gamma_r$. Volumetric recombination is, thus, not strictly required for plasma detachment. Instead, the particle flux can reduce through 'power limitation' - a reduction of ion source induced by a reduction of $q_{recl}$ and/or increase of $E_{ion}$. 


Equation \ref{eq:PowerPartBal} must, however, be consistent with with the marginal Bohm criterion at the plasma sheath ($\Gamma_t \propto p_t/\sqrt{T}$). We can use the sheath target conditions to derive equation \ref{eq:MomBal}, which shows the importance of target pressure losses for reducing the ion target flux \cite{Stangeby2018,Stangeby2017,Verhaegh2019}. For simplicity, electron-ion recombination is ignored in equation \ref{eq:MomBal} (e.g. $\Gamma_t \approx \Gamma_i$ is assumed) \footnote{Recombination could, however, be included by introducing it in an 'effective' $E_{ion} = \frac{E_{ion}}{1-\alpha}$ where $\alpha = \Gamma_r/\Gamma_i$. In that case, recombination can be included in all the ionisation-only formulations of equations \ref{eq:PowerPartBal} and \ref{eq:MomBal} by replacing $E_{ion}$ with $E_{ion}^{eff}$.}.

\begin{equation}
    \Gamma_t = \frac{\gamma p_t^2}{2 m_i q_{recl}} \frac{\frac{\gamma T_t}{E_{ion}}}{1 + \frac{\gamma T_t}{E_{ion}}} = \frac{\gamma p_t^2}{2 m_i q_{t}^{kin}}
    \label{eq:MomBal}
\end{equation}

Equations \ref{eq:PowerPartBal} and \ref{eq:MomBal} may appear to be two different models. However, both can be derived from a model which combines the Bohm criterion with power and particle balance. In that model both formulations are equivalent \cite{Verhaegh2019}: they are thus like two sides of the same coin.  One focuses on power and particle balance while the other upon the momentum balance. Any solution will require all three to balance: consideration of one implies the other one is also true.

Most experimental studies on detachment focus on the macroscopic behaviour of detachment: target heat flux, target temperature, target particle flux, volumetric radiation, etc. Fewer focus upon the underlying atomic (and yet fewer molecular) reactions that generate the necessary particle, power and momentum losses. Plasma spectroscopy in the divertor region can be used to isolate the individual atomic and molecular processes resulting in power and particle losses, providing deeper insight into the detachment process. 

\subsection{Plasma detachment investigations using spectroscopy}

Such studies already commenced a few decades ago \cite{Krasheninnikov1997,Lipschultz1999,Terry1998,Lipschultz1999a,Terry1999,Wenzel1999} by monitoring the Electron-Ion Recombination (EIR) ion sink on various devices (C-Mod, AUG, DIII-D) \cite{Terry1998,Wenzel1999,Isler1997} using spectroscopy of the high-n hydrogen Balmer lines. Although the EIR ion sink was observed to be significant for some detached conditions, it could not always explain the magnitude of the ion target flux loss, particularly for seeded conditions \cite{Lipschultz1999}. This raised the suspicion that the reduction of the ion target flux may result from a reduction of the divertor ion source; which had been hypothesised to occur as the power entering the ionisation region limits that which can be spent on ionisation – power limitation or ‘power starvation’ \cite{Krasheninnikov1997,Lipschultz1999,Pshenov2017,Verhaegh2019}. More recent spectroscopic investigations have directly confirmed a reduction of the divertor ion source (ionisation rate) during detachment \cite{Verhaegh2019,Verhaegh2019a,Lomanowski2019,Lomanowski2020}, which was shown to result from power limitation in \cite{Verhaegh2019} where it occurs simultaneously with volumetric momentum losses. Estimates of the ion source on JET using Ly$\alpha$ intensities \cite{Lomanowski2019,Lomanowski2020} indicated accounting for opacity/photon absorption \cite{Pshenov2019} can be crucial in both the analysis of spectroscopic data and in the assessment of the effective ionisation and recombination cross-sections \cite{Pshenov2019} – with the accompanying reduction in ionisation energy losses - $E_{ion}$. 

The above experimental studies have focused upon the impact of plasma-atom interactions on detachment. Plasma-molecule interactions can also impact power, particle and momentum losses. Such collisions can transfer power and momentum from the plasma to the molecules \cite{Park2018,Moulton2017} and excite the molecules rovibronically, leading to emission in molecular bands such as the Fulcher band ($d \: 3\Pi u \rightarrow a \: 3 \Sigma g$ transition) at 590-640 nm \cite{Fantz2001,Fantz2002,Hollmann2006}. Reactions between the plasma and molecules can produce molecular ions, such as $D_2^+$, which undergo further reactions with the plasma. These ions can result in Molecular Activated Recombination (MAR) or Ionisation (MAI), which act as further ion sinks/sources \cite{Krasheninnikov1997,Terry1998,Fantz2001,Groth2019,Hollmann2006}. Ultimately, plasma-molecule reactions can form excited hydrogen atoms, which emit atomic line emission \cite{Wuenderlich2016,Wunderlich2020} and thus serve as an additional plasma energy loss. The impact of such effects is difficult to study directly but may be extremely important in determining the divertor plasma characteristics.
 
Recent developments in our spectroscopic analysis technique (BaSPMI \cite{Verhaegh2021}) have enabled the inference of the impact of plasma-molecule interactions through their effect on the hydrogenic line series intensities. Initial application of this technique to the TCV tokamak ($n_e \sim 10^{20} m^{-3}$) divertor plasma indicated that plasma-molecule interactions can: 1) have a strong impact on the observed Balmer line series; and 2) result in a large ion sink through MAR, that is significantly larger than electron-ion recombination \cite{Verhaegh2021a}.

\subsection{This paper}

In this paper we build on those previous experimental results \cite{Verhaegh2021a} and use the BaSPMI \cite{Verhaegh2021} technique to investigate plasma detachment in TCV further to provide a more complete view of previous findings. 

New quantitative analysis in this work shows:

\begin{enumerate}
    \item \emph{plasma-molecule interactions} result in \emph{excited atoms} that contribute up to 50 \% of the total \emph{atomic hydrogenic radiative} losses (section \ref{ch:PradPart}). 
    \item indirect experimental evidence for  $D^-$ near the target in the deepest detached conditions (section \ref{ch:detach_dyna}). This suggests a presence of highly vibrationally excited molecules. 
    \item the ion target flux reduction is induced both by an ion sink due to MAR as well as a reduction of the ion source due to 'power limitation' (section \ref{ch:PradPart}).
    \item Molecular Activated Dissociation (MAD) is non-negligible (section \ref{ch:D2p_MAD}) and occurs below the ionisation region, which may play a significant role in power balance. 
    \item $N_2$ seeded discharges (section \ref{ch:N2_MAR}) are lacking MAR as the target temperatures obtained during detachment (3.5-6 eV) are higher than the temperature regime in which MAR is expected to occur (1.5-3 eV), \ref{ch:MolTe}.
\end{enumerate}




Detailed comparisons between the above experimental findings and SOLPS-ITER modelling are presented, which show:
\begin{enumerate}
    \item the impact of $D_2^+$ is strongly underestimated in the SOLPS-ITER simulations (section \ref{ch:SOLPS_MAR}). It is hypothesised this is related to the molecular charge exchange $D_2 + D^+ \rightarrow D_2^+ + D$ rate used in SOLPS-ITER (section \ref{ch:discuss_H2pHm_rate}).
    \item post-processing of the already converged SOLPS-ITER simulations, using a modified molecular charge exchange rate from \cite{Kukushkin2017}, results in a quantitative agreement between simulation and our experimental results (section \ref{ch:SOLPS_MAR}). 
\end{enumerate} 

\section{Experimental results on TCV}
\label{ch:results}

In this work, a density ramp (sections \ref{ch:PradPart}, \ref{ch:detach_dyna}) as well as $N_2$ seeding (section \ref{ch:N2_MAR}) in the TCV tokamak \cite{Coda2019} ($D_2$ fuelling, carbon walls) are used to achieve detachment. All discharges operate in reversed field (i.e. unfavourable to H-mode), employ no auxiliary heating, have an open divertor (no baffles), employ a single null divertor configuration and operate at 340 kA ($q_{95} \approx 2.4$). The measurements and studies discussed in this work are obtained for the outer divertor leg. In these reversed field conditions (before baffles were installed \cite{Reimerdes2021}), the outer target detaches first while the inner target remains attached during the density ramps. During $N_2$ seeding, however, both inner and outer target roll-over together during detachment \cite{Fevrier2019submitted}. Feedback control is used on the $N_2$ seeded discharges to maintain the core electron density stable at a value similar to the detachment onset, which was obtained through a density ramp. The main diagnostic used in this work is the Divertor Spectroscopy System - DSS \cite{Verhaegh2017,Verhaegh2018,Verhaegh2019}.



The spectroscopic analysis in this work is performed using the BaSPMI technique (described in detail elsewhere \cite{Verhaegh2021,Verhaegh2021a,Verhaegh2019}), which uses electron density estimates derived from a Stark broadening analysis in combination with $D_\alpha-D_\gamma$ measurements.  Following previous research \cite{Verhaegh2021,Verhaegh2021a}, the contribution of $D_3^+$ is ignored by BaSPMI. Further information can be found in \ref{ch:BaSPMI}.

\begin{figure}[H]
    \centering
    \includegraphics[width=0.35\linewidth]{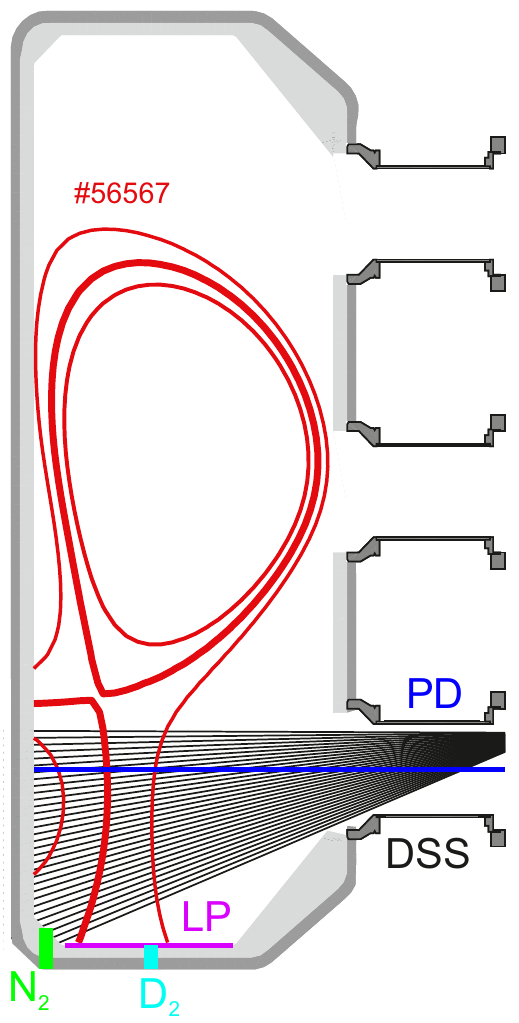}
    \caption{Lines of sight for the DSS (black) as well as photodiode (PD - blue) and plasma geometry for \# 56567, together with Langmuir probe (LP) coverage as well as $D_2$ and $N_2$ fuelling/seeding location.}
    \label{fig:LoS}
\end{figure}

The important diagnostic locations (DSS, Langmuir probes \cite{Fevrier2018,Oliveira2019} and photodiode chord), as well as the magnetic geometry and fuelling / $D_2$ puffing location are shown in figure \ref{fig:LoS}. 

\subsection{Power and particle balance with plasma-molecule interactions during a density ramp}
\label{ch:PradPart}

\begin{figure}[H]
    \centering
    \includegraphics[width=0.55\linewidth]{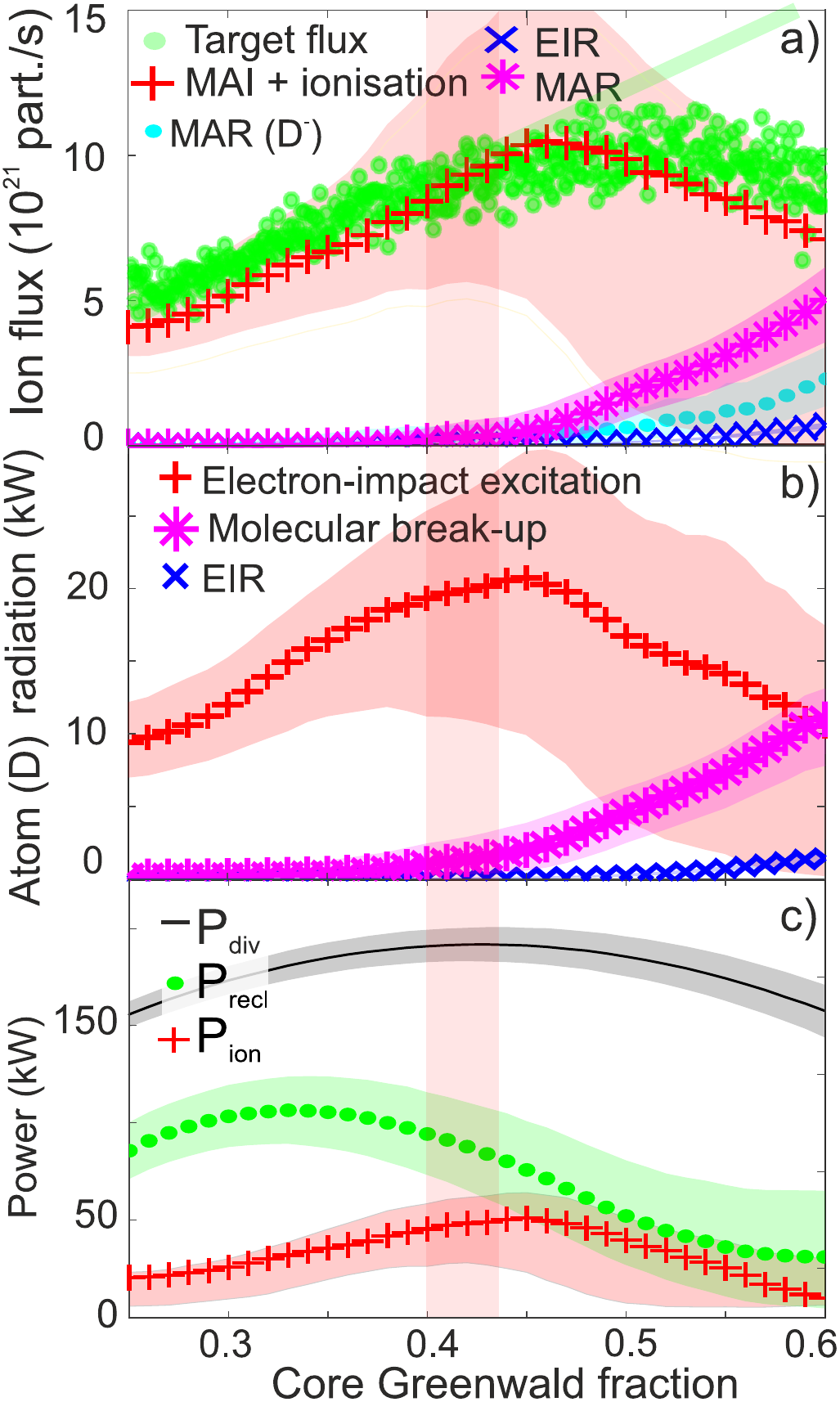}
    \caption{Particle balance (a), hydrogenic radiative losses (b) and power balance (c) as functions of core Greenwald fraction for a density ramp discharge. a) The target flux is measured by Langmuir probes \cite{Fevrier2018,Oliveira2019} (green) with a linear fit from before detachment \cite{Verhaegh2021a,Verhaegh2019}. Spectroscopically inferred atomic and molecular-activated ion source (red) from \cite{Verhaegh2021a}, electron-ion recombination ion sink (EIR - blue) from \cite{Verhaegh2021a}, total molecular activated recombination ion sink (MAR - magenta) from \cite{Verhaegh2021a}, $D^-$ related MAR ion sink (cyan) (\emph{new}) - all integrated over the outer divertor volume. b) \emph{new} - Atomic hydrogen  radiative losses due to electron-impact excitation (red), electron-ion recombination (blue) and molecular break-up from plasma-molecule interactions (magenta). c) \emph{new} - Power entering the outer divertor leg (black - $P_{div}$), power entering the recycling region (green - $P_{recl}$) and power required for ionisation (red - $P_{ion}$). The red vertical shaded region indicates an estimate of the detachment onset; where the measured target flux starts to deviate from a linear increase. The experimental data is from discharge \# 56567 and consecutive repeats.}
    \label{fig:PowerPartBal}
\end{figure}

The particle balance obtained during a core density ramp was analysed previously in \cite{Verhaegh2021a}, which combined contributions of $D_2^+$ and $D^-$ and indicated that a MAR ion sink starts to occur during detachment onset (vertical shaded region) increasing to around 50\% of the measured ion target flux at the end of the density ramp (Fig. \ref{fig:PowerPartBal}). 
Predictions of the $D^-$-only MAR ion sinks are shown separately in Fig. \ref{fig:PowerPartBal}a. Although the $D^-$ creation cross-section is expected to be much smaller than that of $H^-$ \cite{Krishnakumar2011}, our analysis suggests MAR involving $D^-$ may occur, requiring further investigation. This may be attributed to the presence of highly vibrationally-excited molecules, where the isotope difference between creating $D^-$ and $H^-$ is reduced \cite{Fabrikant2002, Reiter2018}; this question is further examined in sections \ref{ch:detach_dyna}, \ref{ch:discuss_H2pHm_rate}.


Previous investigations on DIII-D suggested that the radiative losses from \emph{molecules themselves} through radiative bands of the Werner and Lyman series ($C \: 1\Pi u \rightarrow X \: 1 \Sigma g$ and $B \: 1 \Sigma u \rightarrow X \: 1 \Sigma g$ transitions, respectively) may be unimportant \cite{Groth2019}, as they were not observed. However, molecular break-up from plasma-molecule interactions can result in excited hydrogen atoms and thus radiative losses, which are estimated from our BaSPMI analysis. These occur simultaneously with MAR in figure \ref{fig:PowerPartBal}b (magenta). The resulting hydrogenic radiation can provide up to half of the total hydrogenic radiation. Further power losses may occur from dissociation (section \ref{ch:E_MAR}) as well as energy transfer from the plasma to the molecules through collisions \cite{Myatra,Park2018,Moulton2017}.



Power and particle balance are closely related as it takes power to ionise neutrals, providing the ion source that provides the ion target flux. A certain amount of power, $P_{div}=\int q_{div}$ (equation \ref{eq:PowerBal}), is estimated to enter the outer divertor leg, which is approximately half the power crossing the separatrix into the SOL (assuming a 1:1 outer/inner divertor power symmetry \cite{Verhaegh2019}); $P_{div}$ is estimated in our study from the power due to Ohmic heating minus core radiative losses (assumed to be the sum of the radiative losses above the x-point) \cite{Verhaegh2019}. This power is partially dissipated through impurity radiation ($P_{rad}^{imp}$), estimated by taking the measured total radiation from bolometry (at the outer divertor) minus the inferred hydrogenic radiative losses estimated using spectroscopy. There is a substantial difference between the remainder of the power, $P_{recl} = P_{div} - P_{rad}^{imp} = \int q_{recl}$ (equation \ref{eq:PowerBal}) that enters the recycling region, and $P_{div}$ even in a density ramp discharge as radiative losses from intrinsic carbon are not negligible. 

Fig. \ref{fig:PowerPartBal} shows that, after a Greenwald fraction of 0.35 is reached in the discharge, $P_{recl}$ decreases until it approaches $P_{ion}$ to within a factor 2 at detachment onset ($\sim0.4-0.42$ Greenwald fraction), which is the power required for ionisation, estimated using spectroscopy. At this onset (shaded) the ion target flux starts deviating from its linear increase and $T_t\approx E_{ion}/\gamma \sim 4-6$ eV is reached in agreement with analytical theory \cite{Verhaegh2019} - equation \ref{eq:PowerPartBal}. Around that point in the discharge the divertor ion source no longer increases linearly and ultimately rolls-over ($\sim0.47$ Greenwald fraction), which accounts for a large fraction of the observed ion target flux loss (Fig. \ref{fig:PowerPartBal}a) \cite{Verhaegh2021a}. This experimental evidence suggests that the outer divertor ion source is being limited by the power entering the recycling region: ‘power limitation’. This was already demonstrated in \cite{Verhaegh2019} on the basis of a purely atomic analysis but now, with the inclusion of molecular effects, this picture is re-affirmed. Power limitation occurs simultaneously with the development of target pressure losses both due to volumetric momentum losses and upstream pressure reduction \cite{Verhaegh2019}.

As $P_{ion} / P_{recl}$ increases beyond 0.5, the target temperature ultimately reduces below 3 eV, figure \ref{fig:N2_DenRamp_Compa} d, the temperature regime favourable for MAR and, ultimately, electron-ion recombination ($<1.5$ eV). One difference between the earlier results in \cite{Verhaegh2019} (that did not account for plasma-molecule interactions) and those in Fig. \ref{fig:PowerPartBal}c is that, as $P_{recl}$ and $P_{ion}$ approach each other, significant ion sinks from MAR occur in addition to the ion source reduction. Applying equation \ref{eq:PartBal} to the particle balance in Fig. \ref{fig:PowerPartBal}a suggests that ionisation outside of the outer divertor leg occurs giving a net flow of ions towards the target $\Gamma_u$. That trend is consistent with SOLPS-ITER simulations \cite{Fil2017,Fil2019submitted,Wensing2019}, which will be further discussed in \ref{ch:SOLPS_IUp}.

\subsection{Profile evolution of ion/power sources/sinks during detachment evolution} 
\label{ch:detach_dyna}

\begin{figure}[H]
    \centering
    \includegraphics[width=0.9\linewidth]{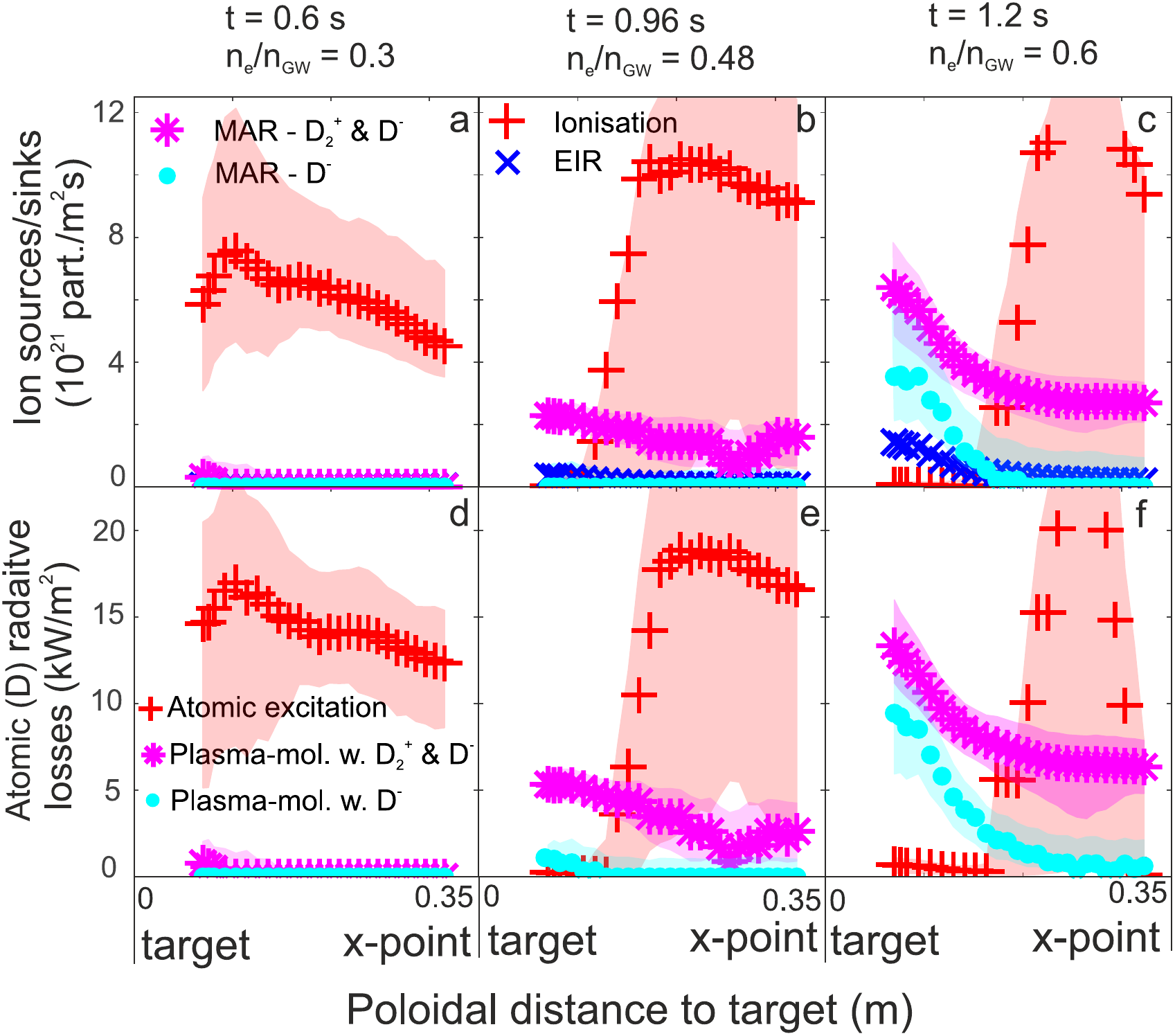}
    \caption{Inferred line-integrated profiles along the divertor leg of ion source/sink (top row, a,b,c) and power loss profiles (bottom row: d,e,f) due to plasma-atom and plasma-molecule interactions for the attached, detachment onset and detached phases. a,b,c) estimated total (atomic and molecular activated) ionisation source; Electron-Ion Recombination (EIR) sink; Molecular Activated Recombination ion sink - MAR - $D_2^+ \: \& \: D^-$; MAR - $D^-$ alone. d,e,f) Atomic radiation due to atomic excitation; radiative losses through \emph{excited atoms arising from plasma-molecule interactions} involving $D_2^+ \: \& \: D^- \: \& \: D_2$ (the latter one is negligible) as well as $D^-$ alone. The experimental data is from discharge \# 56567 and consecutive repeats.}
    \label{fig:PowerPartBalProfs}
\end{figure}

The power and particle sinks/sources discussed previously have been integrated along the outer divertor leg. Here we study the evolution of these interaction profiles quantitatively during detachment (figure \ref{fig:PowerPartBalProfs}), which are compared against SOLPS-ITER simulations in section \ref{ch:SOLPS_Dyna}.  The particle/power sink/source profiles are obtained from \emph{line-integrated estimates} along the lines of sight indicated in figure \ref{fig:LoS}.

The first observation drawn from figure \ref{fig:PowerPartBalProfs} is that the profile for each relevant ion source/sink (first row of figure \ref{fig:PowerPartBalProfs}) and its evolution is very similar to the corresponding hydrogenic radiative loss (second row). A second observation is that MAR attributed to $D^-$ predominantly occurs in the deepest detached conditions close to the target.

As $D^-$ preferentially populates the $n=3$ excited level \cite{Wunderlich2020,Wuenderlich2016,Fantz2006}, the BaSPMI technique estimates the $D^-$ contribution from the ratio of the molecular contribution of $D_\beta$ to the molecular contribution of $D_\alpha$. To gain more insight into the experimental evidence for the presence of $D^-$, we performed an alternative analysis using only $D_{\alpha}, D_{\gamma}, D_{\delta}$ measurements whilst assuming that all plasma-molecule interaction related emission arises from $D_2^+$ (\ref{ch:BaSPMI}). This alternative analysis leads to similar MAR ion sink estimates as the full analysis. However, extrapolating the alternative analysis outputs to $D_\beta$ and comparing this to $D_\beta$ measurements not included in the analysis, shows the alternative analysis overpredicts $D_\beta$ by $\sim 30$ \%. Whereas, the full analysis does match the $D_\beta$ (and $D_\alpha, D_\gamma, D_\delta$) measurements by attributing Balmer line emission to $D^-$. 

The MAR $D^-$ to total MAR ratios are expected to increase at low $T_e$: using AMJUEL data tables \cite{AMJUEL} together with the isotope rescaling coefficients from \cite{Reiter2018} (see \ref{ch:PostProcessing}) the MAR $D^-$ to total MAR ratios are 0.3, 0.17, 0.07 at 1, 2, 3 eV. This strong increase of the contribution of MAR from $D^-$ at low temperatures is qualitatively in agreement with our observations. However, the experimentally inferred ratios of MAR from $D^-$ to total MAR (figure \ref{fig:PowerPartBalProfs}) are higher than the predicted coefficients. This may imply that there is a larger concentration of highly vibrationally-excited molecules than obtained by the vibrational model used to determine the rates used by SOLPS-ITER \cite{AMJUEL,Kotov2007} (see sections \ref{ch:SOLPS_Dyna}, \ref{ch:discuss_H2pHm_rate}). 

\subsection{Comparison between density ramp experiments and $N_2$ seeded discharges}
\label{ch:N2_MAR}

\begin{figure}[H]
    \centering
    \includegraphics[width=0.8\linewidth]{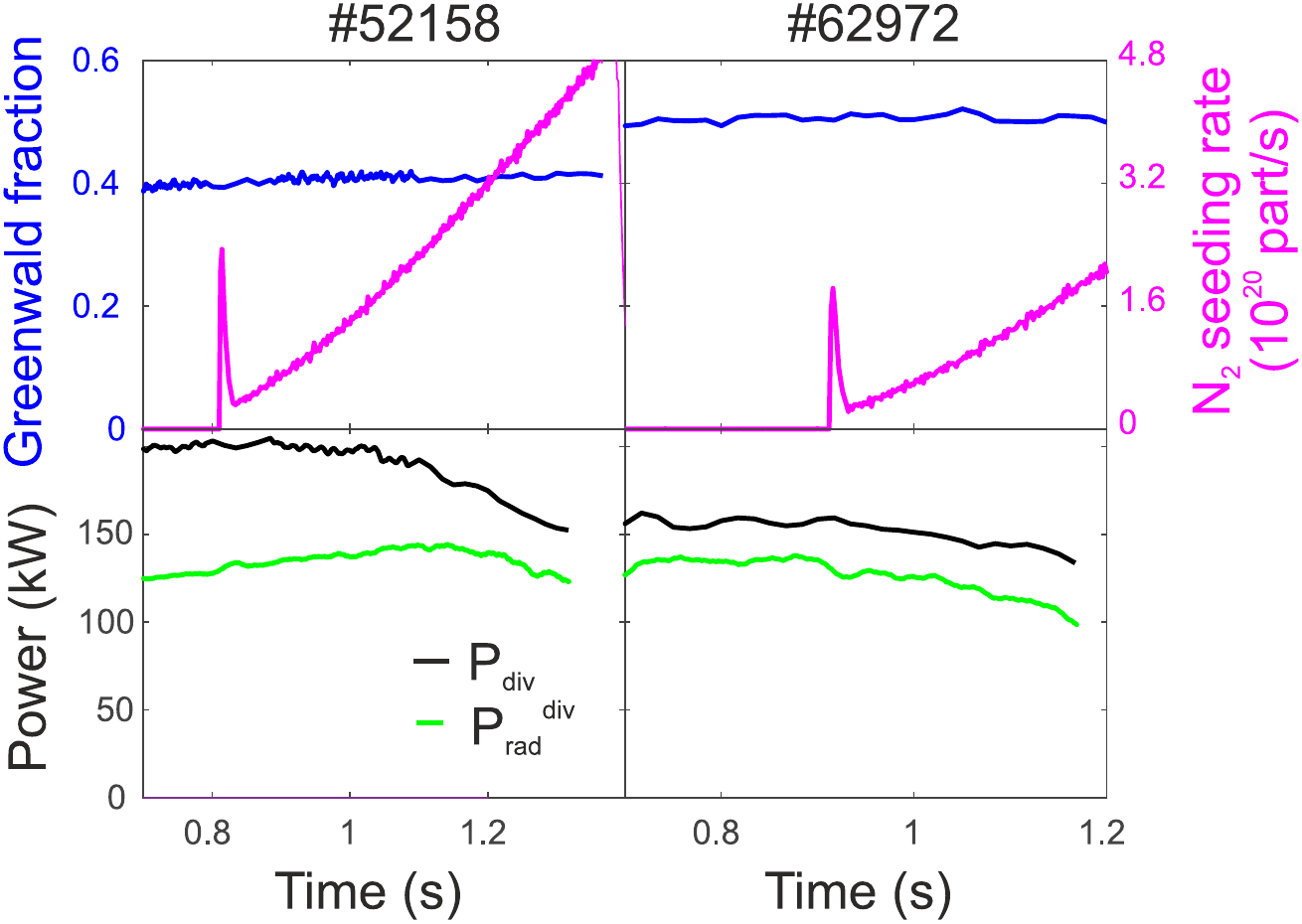}
    \caption{Overview of two $N_2$ seeded discharges in terms of core Greenwald fraction, $N_2$ seeding level, power entering outer divertor $P_{div}$ and outer divertor radiation $P_{rad}^{div}$.}
    \label{fig:N2_Overview}
\end{figure}

In this section we discuss two $N_2$ seeded discharges, for which an overview description is shown in figure \ref{fig:N2_Overview}. The core density is set to a level such that the seeding ramp occurs just \emph{before} the detachment onset for pulse \#52158 and just \emph{after} the detachment onset for pulse \#62972. Plasma-molecule interactions which involves $NH_x$ can occur in $N_2$ seeded plasmas, which is further discussed in \ref{ch:SOLPS_N2} but is not considered below. Such reactions may, however, also result in excited hydrogen atoms and impact the hydrogen Balmer spectra, which requires further study.

Discharge \# 62972 corresponds to a set of repeat discharges such that spectroscopic data was obtained with several different DSS settings to obtain measurements of $D_\alpha$, $D_\delta$ and $D_\epsilon$. In \# 52158, the impact of $N_2$ seeding in reducing the ion target flux is stronger and the difference between the power entering the outer divertor $P_{div}$ and the divertor radiative power losses $P_{rad}^{div}$ is larger before the start of seeding (figure \ref{fig:N2_Overview}). However, only DSS measurements of $D_\delta$ and $D_\epsilon$ were available and to monitor $D_\alpha$  a single line of sight with a non-calibrated filtered photodiode was used (line of sight indicated in figure \ref{fig:LoS}). \footnote{Its measurement (in V) has been rescaled to the extrapolated $D_\alpha$ brightness from analysing the $n=6,7$ Balmer line measurements, from a single corresponding DSS line of sight, with the atomic part of BaSPMI.} For both discharges, a significant amount of residual nitrogen (5-10\% \cite{GahlePSI}) was already present before $N_2$ seeding from previous discharges, increasing to 10-15 \% during $N_2$ seeding. 

\begin{figure}[H]
    \centering
    \includegraphics[width=\linewidth]{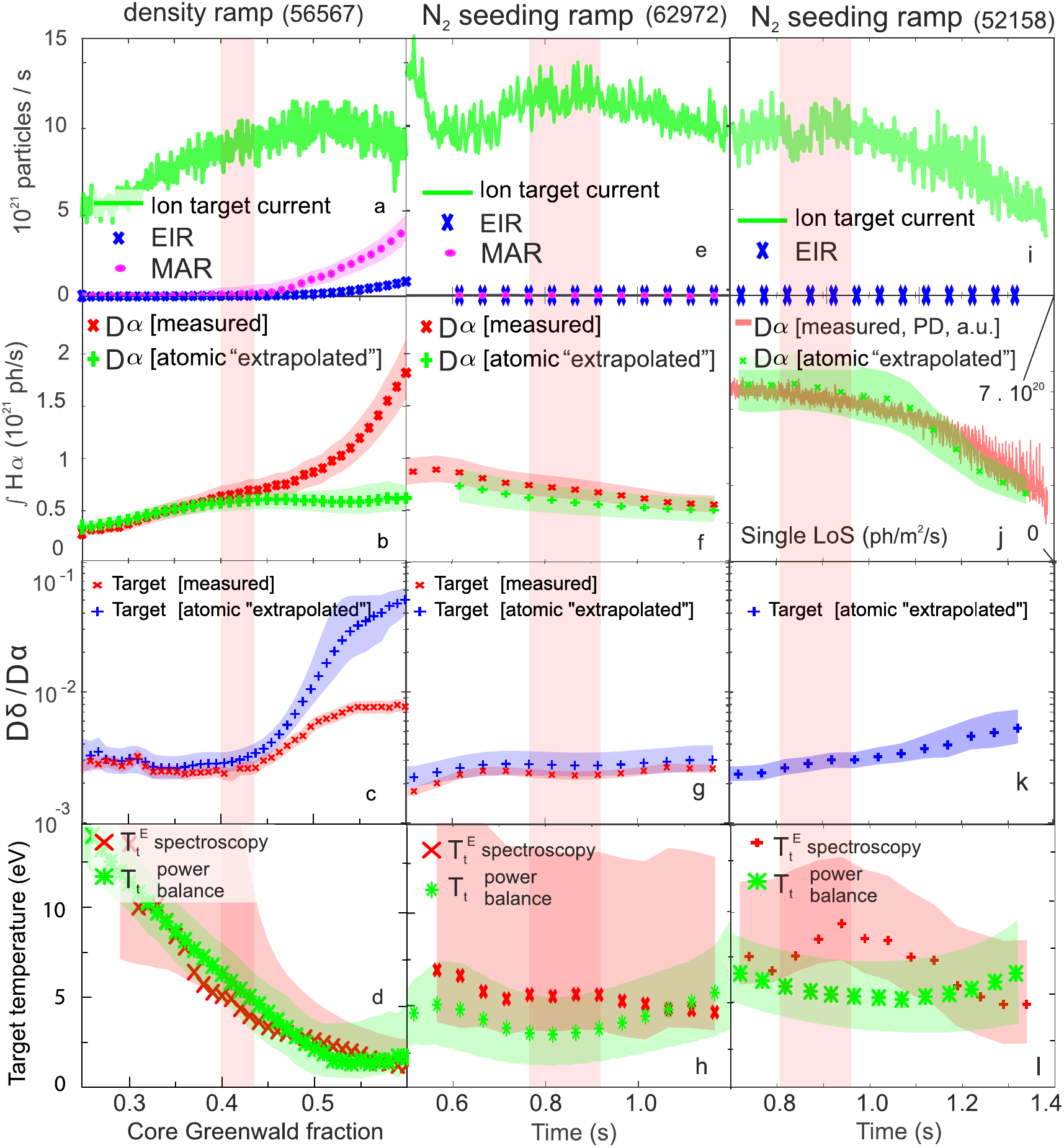}
    \caption{Comparison plot between detachment through a core density ramp (a,b,c,d) and two $N_2$ seeded discharges (e,f,g,h and i,j,k,l). a,e,i) outer target ion flux, electron-ion recombination (EIR) and Molecular Activated Recombination (MAR) (a,e) ion sinks. b,f,j) Measured and 'atomic "extrapolated"' $D_\alpha$ by analysing the higher-n states with the atomic part of BaSPMI. These values are either integrated along the outer divertor leg (b,f) or taken from a single chord (j). c,g,k) Measured and atomic extrapolated $D_\delta/D_\alpha$ line ratios near the target. d,h,l) Estimated target temperatures from a spectroscopic chord closest to the target and from power balance. The y-axis scales are the same for the $N_2$ seeding and core density ramp columns (except for j). Estimates of the detachment onsets are shaded vertically in red.}
    \label{fig:N2_DenRamp_Compa}
\end{figure}

Figure \ref{fig:N2_DenRamp_Compa} shows a comparison of spectroscopically-derived quantities and measurements between discharges that reach detachment through a core density ramp and discharges that are detached through $N_2$-seeding. To keep the analysis consistent between all discharges, only the atomic part of BaSPMI was utilised. Combining this with $D_\alpha$ measurements facilitated estimating EIR and MAR ion sinks \cite{Verhaegh2021}, which are shown in figure \ref{fig:N2_DenRamp_Compa} a,e.  

Our first observation is that $D_\alpha$ drops as the ion target flux rolls-over in the seeded cases (Fig. \ref{fig:N2_DenRamp_Compa} f,j) – consistent with the $D_\alpha$ intensity extrapolated from the purely atomic analysis of the $n=5,6,7$ Balmer lines. This contrasts with the density ramp case where the measured $D_\alpha$ intensity increases during detachment and deviates from the $D_\alpha$ intensity extrapolated from the atomic analysis of the medium-n Balmer lines (Fig. \ref{fig:N2_DenRamp_Compa} b), indicating the presence of plasma-molecule interactions resulting in MAR (Fig. \ref{fig:N2_DenRamp_Compa} a). The absence of this behaviour during the $N_2$ seeded case suggests that these plasma-molecule interactions resulting in MAR are, just as EIR, negligible during $N_2$ seeding (Fig. \ref{fig:N2_DenRamp_Compa} e,i).

A second observation is that the trend in both the monitored and atomic extrapolated $D_\delta/D_\alpha$ ratio (figure \ref{fig:N2_DenRamp_Compa} c,g,k) are different between the $N_2$ seeded and density ramped discharges. During $N_2$ seeding, the atomic extrapolated and the measured line ratios remain close to the value expected for an electron-impact excitation (EIE) dominated plasma ($0.002-0.003$) (figure \ref{fig:N2_DenRamp_Compa} g,k). This indicates EIR does not (or barely for \# 62972) contribute to the medium-n Balmer line emission. In contrast, during the density ramp, the atomic extrapolated $D_\delta/D_\alpha$ ratio strongly rises from $\sim 0.0025$ to $\sim 0.05$ (figure \ref{fig:N2_DenRamp_Compa} c) during detachment as electron-ion recombination strongly contributes to the medium-n Balmer line emission ($> 70 \%$ for n=5,6 \cite{Verhaegh2021}). The measured $D_\delta/D_\alpha$ ratio is smaller ($\sim 0.008$), however, due to plasma-molecule interactions contributions to $D_\alpha$ (figure \ref{fig:N2_DenRamp_Compa} c).

The absence of EIR, MAR and enhancement of $D_\alpha$ in the $N_2$ seeded case is consistent with the measured target temperature (Fig. \ref{fig:N2_DenRamp_Compa} h, l), of around 3-6 eV. This is estimated using the excitation emission from the line of sight closest to the target ($T_t^E$ spectroscopy) and, separately, from the power balance \cite{Verhaegh2019}. Although some of the temperature estimations indicate a significantly higher upper uncertainty bound than 6 eV, it is unlikely the plasma temperature is above 6 eV based on other measurements (Langmuir probes, impurity emission regions).  For the $N_2$ seeded conditions, the electron temperature (within uncertainties) as well as the inferred electron density ($\sim 1-3 \times 10^{20} \textrm{m}^{-3}$) is constant along the divertor leg.  Meanwhile, the temperature is observed to decrease during the core density ramp discharge (Fig. \ref{fig:N2_DenRamp_Compa} d), reaching 3 eV at the MAR onset and 1 eV at the end of the discharge. Therefore, the $N_2$ seeded conditions do not reach the temperatures at which there is a sufficiently high molecular density for MAR to play a strong role, which is supported by the more detailed analysis of MAR's $T_e$ regime in \ref{ch:MolTe}. This indicates that plasma-molecule interactions, resulting in MAR, do not necessarily dominate in all detached conditions. 

\section{Comparisons against SOLPS-ITER simulations}

A detailed comparison between SOLPS-ITER simulations of TCV detachment through core density ramp (from \cite{Fil2017,SOLPS_Fil} using synthetic diagnostics) and experimental spectroscopic data (analysed from an atom-interaction only perspective) previously showed good agreement, in general \cite{Verhaegh2019}. 

The simulations were performed using SOLPS-ITER version 3.0.6-57-g7705c5 \cite{Wiesen2015}. The recycling coefficient is set to 0.99 for all surfaces (e.g. wall pumping) to match both the ion target flux profiles as well as the fuelling levels. Carbon chemical sputtering was set to 3.5 \% to match the measured CIII (465 nm) brightnesses. All puffed particles are initially molecules. Additional molecules can be generated upon reflection, according to the TRIM database \cite{Reiter2005} for deuterium reflecting on carbon. Vibrational states were not explicitly included in the SOLPS-ITER simulation (but are instead included in the effective reaction rates (see section \ref{ch:SOLPS_MAR})). This implies that the impact of the plasma-wall interaction on the vibrational levels is not accounted for in the SOLPS-ITER simulations.

\subsection{MAR and $D_\alpha$}
\label{ch:SOLPS_MAR}

\begin{figure}[H]
    \centering
    \includegraphics[width=\linewidth]{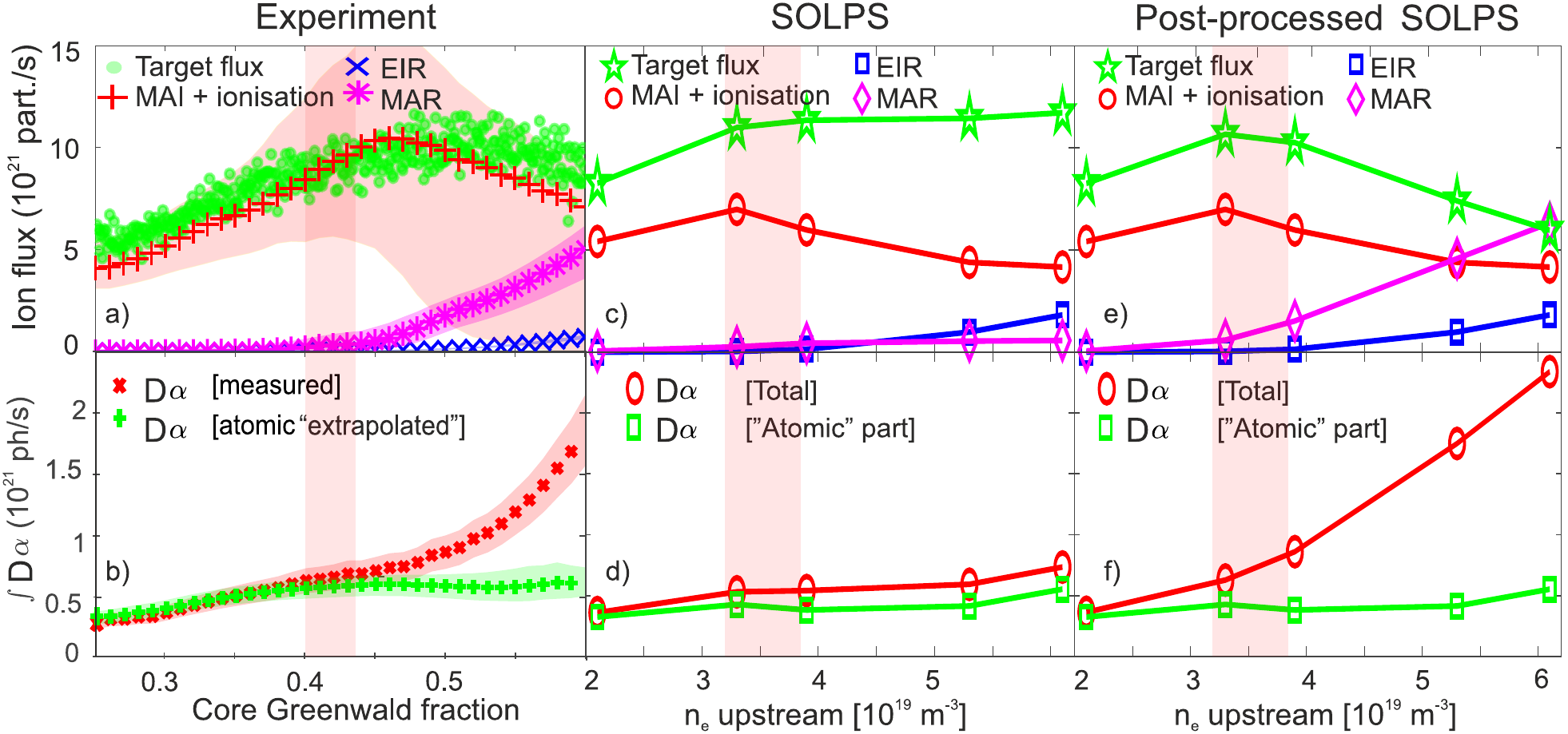}
    \caption{Particle balance and measured/atomic ‘extrapolated’ $D_\alpha$ in the outer divertor (integrated) from measurements and BaSPMI analysis (a,b), compared against that obtained using a synthetic divertor spectrometer in SOLPS-ITER simulations (c,d) \cite{Fil2017}. The last column plots (e,f) are obtained by post-processing SOLPS-ITER using molecular charge exchange ($D^+ + D_2 \rightarrow D + D_2^+$) obtained from \cite{Kukushkin2017} (\ref{ch:PostProcessing}). Estimates of the detachment onset are shown with vertical shaded red regions. The experimental data is from discharge \# 56567 and consecutive repeats.}
    \label{fig:Compa_Simulation_Integrated}
\end{figure}

A disagreement between experiment and simulation was, however, observed in the $D_\alpha$ brightness \cite{Verhaegh2018} (Fig. \ref{fig:Compa_Simulation_Integrated} b vs d) and in the ion target flux  (Fig. \ref{fig:Compa_Simulation_Integrated} a vs c), which does not roll-over in the simulations as an additional flow of ions ($\Gamma_u$) towards the target compensates the reduction of ionisation in the outer divertor leg \cite{Verhaegh2019,Fil2017,Fil2019submitted,Wensing2019}, generated mainly by ionisation upstream \cite{Fil2019submitted,Fil2017}.

Fig. \ref{fig:Compa_Simulation_Integrated} a,b,c,d illustrates this disagreement in $D_\alpha$ brightness, supporting the idea that the disagreement is caused by the plasma-molecule contribution to $D_\alpha$ being underestimated by the simulation. Likewise, the MAR rate (from $D_2^+$) appears strongly underestimated in the simulation.

These SOLPS-ITER simulations used the default reaction sets \cite{Fil2017} which are effective (grouped) rates: the individual vibrational states ($\nu$) of $D_2$ in the plasma are not individually traced. Instead, the distribution of vibrational states ($f_{\nu} (T_e, T_i)$) is modelled, assuming the vibrational states are in equilibrium \cite{Kotov2007}, using electron and ion impact vibrational excitation, electron impact de-excitation as well as vibrationally resolved charge exchange, ionisation, dissociation and dissociative attachment depletion channels of the individual vibrational states \cite{Kukushkin2017, Reiter2018,Kotov2007,Greenland2001,Reiter2008,AMJUEL}. As this contains both electron and ion interactions, it is both dependent on $T_e$ and $T_i$. The effective rates are then computed as $<\sigma v>_{eff} = \sum_{\nu} f_{\nu} (T_e, T_i) <\sigma v>_{\nu} (T_e, T_i)$ . This is further simplified assuming $T=T_i=T_e$.

Apart from the isotope dependencies of the various rates, the ion temperature dependent rates are \emph{actually} dependent on the relative velocity ($<\sigma v>_\nu (v_{rel})$). \emph{However}, at the same ion temperature, the relative velocity $v_{rel}$ is different because of a mass effect \cite{Kotov2007,Reiter2008}. Because of this, EIRENE rescales, by default, the effective 'grouped' molecular ion temperature dependent rates: $<\sigma v>_{eff} (T_{H_2^1}) = <\sigma v>_{eff} (T_{H_1^1}/2)$ \cite{Kotov2007}.

Our hypothesis is that an underestimation of the impact of $D_2^+$ arises from the above rescaling applied to the molecular charge exchange rate ($D^+ + D_2 \rightarrow D + D_2^+$). As shown, the mass rescaling of the effective rates used also implies that the \emph{electron temperature dependency} of $f_{\nu} (T_e,T_i)$ is rescaled, which we feel is oversimplified. In \cite{Kukushkin2017} the mass rescaling is only applied to the ion temperature dependencies. However, even $<\sigma v>_\nu (v_{rel})$ can have isotope dependencies due to the different molecular structures of the isotopes. In \cite{Reiter2018}, such isotope dependencies are considered for $H^+ + H_2 \rightarrow H_2^+ + H$ and $e^- + H_2 \rightarrow H_2^- \rightarrow H^- + H$, whilst $f_\nu$ is simplified with an isotope independent Boltzmann distribution \footnote{This is different from the $f_{\nu} (T_e, T_i)$ used by SOLPS-ITER. However, more detailed analysis in \cite{Laporta2021} indicates that there may not be a strong deviation of $f_\nu$ from a Boltzmann distribution}. Both these approaches lead to rates which are much closer to the hydrogen rate, which is further discussed in section \ref{ch:discuss_H2pHm_rate} and \ref{ch:PostProcessing}.


To investigate the possible impact these different rates have on SOLPS-ITER simulations of TCV divertor detachment, the converged simulation results were post-processed using the $D^+ + D_2 \rightarrow D + D_2^+$ rate from \cite{Kukushkin2017} to calculate new $D_2^+/D_2$ ratios (\ref{ch:PostProcessing}). Caveat: since this is post-processing, the result is no longer a self-consistent SOLPS-ITER solution as a change in the $D_2^+/D_2$ ratio could change the plasma solution. 

Post-processing, when either the molecular charge exchange rates for hydrogen (H) from \cite{AMJUEL} or the deuterium rates from \cite{Kukushkin2017} or the D to H ratios from \cite{Reiter2018} with the H rates from \cite{AMJUEL} are used, results in significant plasma-molecule interaction contributions to $D_\alpha$ as well as substantial MAR rates. This seems to result in a closer quantitative agreement between the experiment and simulation. \footnote{Studies with filtered camera systems \cite{Perek2021,Harrison2017} have identified the measured $D_\alpha$ emission region near the target extends further into the private flux region than predicted by plasma-edge simulations. Although this effect may explain underestimations in the simulated $D_\alpha$ brightness, compared to the experiment, this cannot explain the relative lack of plasma-molecule interaction contributions in the simulation.}. Additionally, the ion target flux in the post-processed simulations (Fig. 5e) now, indeed, rolls over as the additional influx of ions from outside the divertor leg ($I_u$) during detachment is partially counteracted by the increase in total ion sink ($I_r$) due to the increase in MAR (particle balance - equation \ref{eq:PartBal}).


\subsection{Ion source/sink profile evolution during detachment}
\label{ch:SOLPS_Dyna}



\begin{figure}[H]
    \centering
    \includegraphics[width=\linewidth]{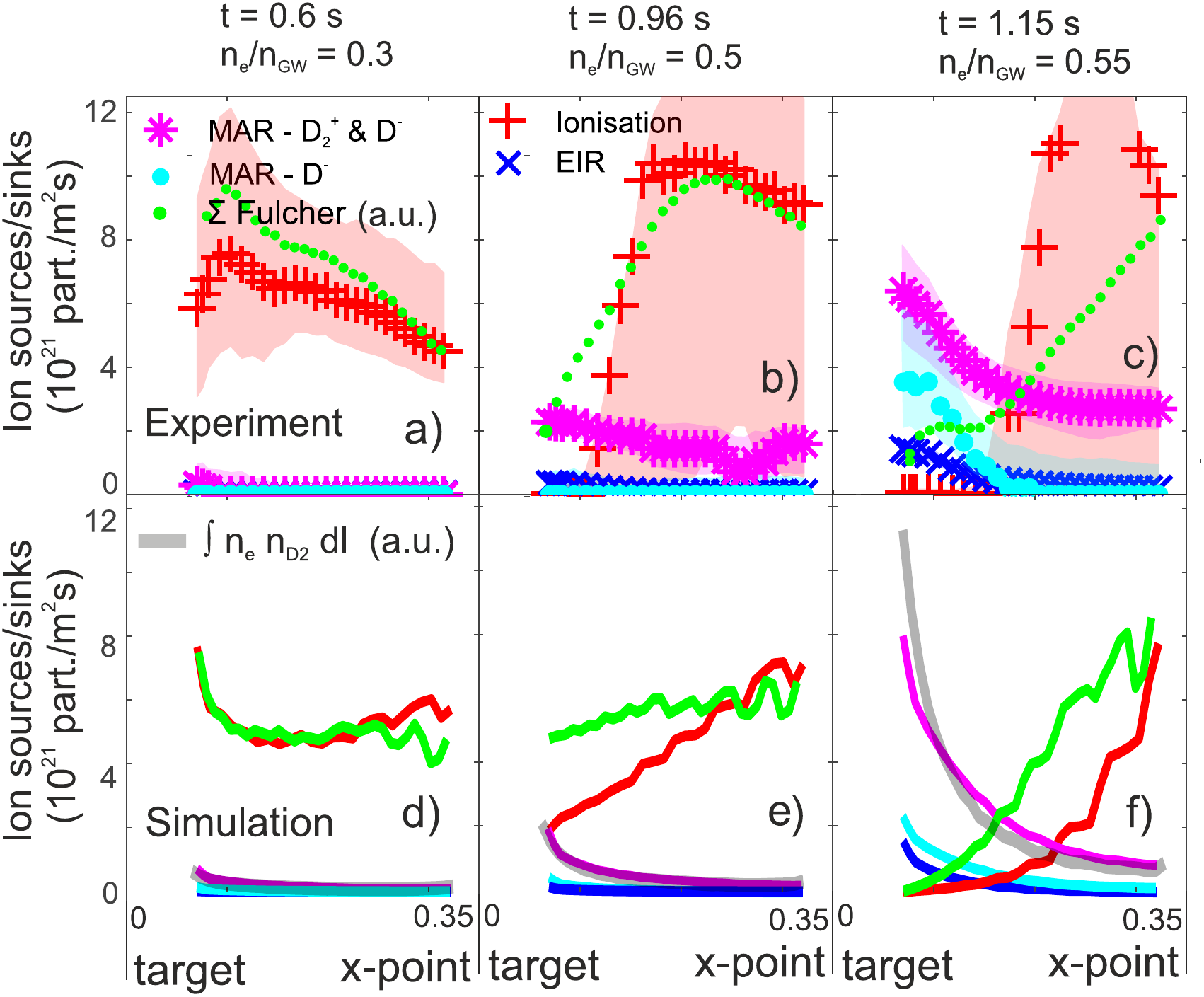}
    \caption{Comparison of ion source/sink profiles along the outer divertor leg between experiment (a,b,c - copied from figure \ref{ch:detach_dyna} a,b,c) and simulation (d,e,f) at three different times/densities of ionisation (red), EIR (blue), MAR ($D_2^+ \: \& \: D^-$) (magenta), MAR $D^-$ (cyan) as well as Fulcher band emission (green). The SOLPS-ITER synthetic diagnostic results also show $\int n_e n_{D_2} dl$ along the line of sight (black - shaded) in arbitrary units. The used lines of sight are indicated in figure \ref{fig:LoS}. In the experiment, the Fulcher band is summed over 600 to 614 nm and presented in arbitrary units, whereas for the simulation,  relative values for the Fulcher band brightness were obtained from AMJUEL rate H12 2.2.5fl \cite{AMJUEL}.  The experimental data is from discharge \# 56567 and consecutive repeats.}
    \label{fig:Compa_Simulation_LoS}
\end{figure}

In this section we compare the evolution of detachment between simulation and experiment, which was discussed in section \ref{ch:detach_dyna}. The SOLPS-ITER simulations were matched to the experimental times through the estimated upstream density, using linear interpolation (before the roll-over) and extrapolating that trend after the roll-over. This was required as an ordering parameter as the upstream density rolls-over/saturates during detachment and follows a previous approach  \cite{Verhaegh2019}. 

The simulated behaviour, after post-processing, is in quantitative agreement with the experiment in most cases. In the attached phase ($t = 0.6$ s, $n_e/n_{GW} = 0.3$), MAR is negligible and ionisation is spread over the entire outer divertor leg. At the detachment onset ($t=0.96$ s, $n_e/n_{GW} = 0.5$), MAR commences near the target and the experimental ionisation profile detaches from the target showing a pronounced ionisation front, whereas the simulated profile appears more smooth. Into the detached phase ($t=1.15$ s, $n_e/n_{GW} = 0.55$), the ionisation front moves further towards the x-point, MAR increases near the target and EIR together with MAR that is attributed to $D^-$, starts to occur near the target. The ratio between MAR from $D^-$ to the total MAR obtained from BaSPMI analysis is higher than the one obtained from post-processed SOLPS-ITER simulations, which is consistent with the findings in section \ref{ch:detach_dyna}. 


Both Balmer line emission from plasma-molecule interactions as well as Fulcher band emission depends on electrons (or ions) interacting with molecules. Therefore, we also show profiles of the integral of the electron density times the deuterium molecule density along the line of sights for the SOLPS-ITER modelling case (grey curves figure \ref{fig:Compa_Simulation_LoS} d,e,f). This indicates a similar trend as the Balmer line emission arising from interactions with $D_2^+$ and has a very different behaviour than the Fulcher band brightness. Therefore, the Fulcher band brightness trend seems to be mostly determined by the electron temperature, rather than the molecule and electron densities, which is further discussed in section \ref{ch:D2p_MAD}. This complicates estimating MAR or the molecular density near the target using Fulcher band emission measurements in strongly detached conditions as the $D_2$ density/Fulcher photons ratio is extremely sensitive to $T_e$. 

\subsection{SOLPS-ITER simulations and $N_2$ seeding}
\label{ch:SOLPS_N2}

Post-processing $N_2$ seeded SOLPS-ITER simulations for TCV \cite{Smolders} in the same way results in a negligible impact of $D_2^+$ in contrast to the impact during a core density ramp because the target temperature reach a minimum near 3.5 eV during these simulations. These higher target temperatures lead to a lower $D_2$ density and thus a lack of $D_2^+$, even when post-processing is applied. Those simulation results are qualitatively consistent with the experimental findings presented in section \ref{ch:N2_MAR}. However, volumetric momentum losses did not occur in these simulations, in contrast to $N_2$ seeding experiments \cite{Fevrier2019submitted}.

Nevertheless, qualitatively the same behaviour (higher $T_t$ during $N_2$ seeding compared to a density ramp) is observed in MAST-U simulations \cite{Myatra} where significant volumetric momentum losses do occur. In these cases, the impact of $D_2^+$ is also small after post-processing \cite{Verhaegh2021}. 

The plasma can also undergo chemical reactions with nitrogen, resulting in the formation of $NH_x$, which can also lead to molecular activated recombination (N-MAR) \cite{Perillo2018,Ezumi2019}. Such reactions are not present in the current SOLPS-ITER simulations and the observed reduction of the ion target flux is attributed to a reduction of the ion source due to power limitation in the simulations, consistent with experimental findings \cite{Verhaegh2019}. Therefore, the relatively high target temperatures during $N_2$ seeded detachment may be caused by something else than N-MAR. One hypothesis is that this is induced by a reduction of neutral (atom and molecule) densities in the divertor compared to a density ramp, resulting in a a different $f_{mom} (T_t)$ trend \cite{Stangeby2017,Verhaegh2019,Myatra}. The impact of N-MAR requires further study, for instance by estimating the N-MAR ion sink experimentally and comparing it to the ion source. Detailed nitrogen spectroscopy comparisons between SOLPS-ITER and TCV indicates a mismatch in the transport of lower nitrogen charge states, which may be attributed to a lack of nitrogen chemistry in the SOLPS-ITER simulations \cite{GahlePSI}.

\section{Discussion}
\label{ch:discussion}

\subsection{$D_2^+$ and $D^-$ creation rates}
\label{ch:discuss_H2pHm_rate}

There is discussion in literature regarding the molecular charge exchange rate which results in $D_2^+$ formation ($D_2 + D^+ \rightarrow D_2^+ + D$) as well as the likelihood of creating $D^-$ ($e^- + D_2 \rightarrow D_2^- \rightarrow D^- + D$) \cite{Kukushkin2017,Reiter2018,Ichihara2000}. There are three important sources of uncertainty for these reaction rates. 

\begin{enumerate}
    \item The reaction rates are sensitive to the vibrational distribution $f_\nu$ \cite{Fabrikant2002,Reiter2018,Ichihara2000}. However, SOLPS-ITER employs effective rates without tracking the individual vibrational levels. This is an important concern as it leads to a large (un-estimated) uncertainty in SOLPS-ITER predictions, even when the isotope dependency of $<\sigma v>_\nu$ is precisely known. 
    \item The reaction rates of ion molecule interactions depend on the relative velocity, which is different at the same temperature for different isotopes. The sequence in which this mass rescaling is applied is important as discussed in section \ref{ch:SOLPS_MAR}.
    \item There can be chemical differences which alter $f_\nu$ or $<\sigma v>_\nu$ between the different isotopologues. These chemical effects lower the cross-section for generating $H^-$ \cite{Krishnakumar2011} at low vibrational levels and increase the vibrational level for which molecular charge exchange becomes resonant \cite{Ichihara2000} for increasing isotope number. Therefore, the exact distribution over the vibrational states will affect the resulting isotope dependency of the reaction rates. 
\end{enumerate}



%


\begin{figure}[H]
    \centering
    \includegraphics[width=0.8\linewidth]{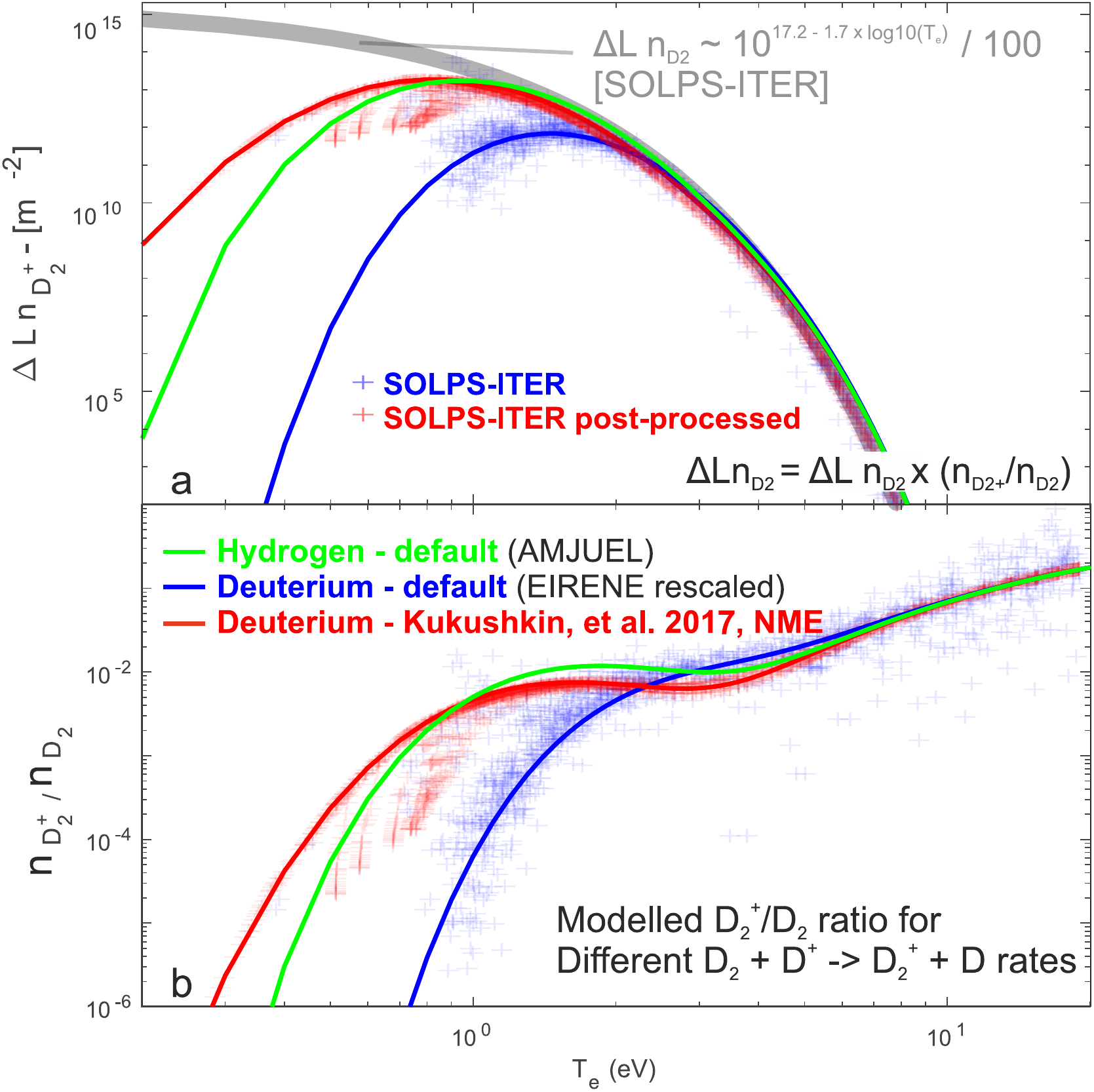}
    \caption{Overview of how different models for molecular charge exchange impact the $D_2^+/D_2$ ratio (b) and the $D_2^+$ density (a). These model curves (solid lines) are weakly sensitive to $n_e$ and $n_e = 5 \times 10^{19} m^{-3}$ is assumed. The impact of those ratios on the $D_2^+$ density is qualitatively estimated by multiplying the $D_2^+/D_2$ ratios with a scaling law for the pathlength times the $D_2$ density obtained from these SOLPS-ITER simulations in \cite{Verhaegh2021} using synthetic diagnostics. Values for $D_2^+/D_2$ for the second most detached SOLPS-ITER case (figure \ref{fig:Compa_Simulation_Integrated}) are shown (blue '+' symbols) as well as their post-processed estimates (red '+' symbols). Small deviations exist for the post-processed case from the red curve as there is a $n_e$ range in the simulation as well as a range of $E_{H_2}$ is assumed which is used in the vibrational distribution calculation \cite{Kukushkin2017}. The shown labels in figure a+b apply to both plots.}
    \label{fig:D2p_Rates}
\end{figure}

Given these caveats, we compare in figure \ref{fig:D2p_Rates}b the modelled $D_2^+/D_2$ density ratio using various molecular charge exchange rates (solid curves) (see \ref{ch:PostProcessing}) against that from SOLPS ITER simulations (blue '+') and post-processing (red '+'). Although the various molecular charge exchange rates are still under discussion, we find that the rates from \cite{Kukushkin2017} (red - Fig. \ref{fig:D2p_Rates}) as well as \cite{Reiter2018} are much closer to the default hydrogen rate (green  - Fig. \ref{fig:D2p_Rates}) than the rescaled deuterium rate used in SOLPS-ITER (EIRENE) (blue - Fig. \ref{fig:D2p_Rates}). This difference is larger at $T_e < 4$ eV, where the molecular density strongly increases with decreasing temperature \cite{Verhaegh2021,Stangeby2017}. This is illustrated in figure \ref{fig:D2p_Rates}a where these $D_2^+/D_2$ ratios are multiplied with a scaling law for the pathlength times the $D_2$ density obtained from SOLPS-ITER simulations of TCV \cite{Verhaegh2021}. This indicates that the peak $D_2^+$ density is $\sim \times 10$ higher when using the default hydrogen rate or the deuterium rates from \cite{Kukushkin2017} and \cite{Reiter2018}, than that for the default EIRENE-rescaled deuterium rate. This explains the strong difference observed, both in terms of MAR and the modelled $D_\alpha$ brightness, between the default SOLPS-ITER and that following post-processing (figure \ref{fig:Compa_Simulation_Integrated}).

\subsection{Molecular Activated Dissociation - MAD}
\label{ch:D2p_MAD}

In this work, we have mainly addressed ion sources/sinks (MAI/MAR) and the hydrogenic radiative losses associated with excited atoms due to plasma-molecule interactions.
However, those reactions are only a small subset of the reactions in which $D_2^+$ and $D^-$ are involved. Several of these reactions lead to the dissociation of molecules - Molecular Activated Dissociation - MAD \cite{Krasheninnikov1997}; the variety of reactions are listed in table 1 in \cite{Verhaegh2021}. MAD can result in an important source of neutral $D$ atoms and plays a role in the power balance (section \ref{ch:E_MAR}). 

\begin{figure}[H]
    \centering
    \includegraphics[width=\linewidth]{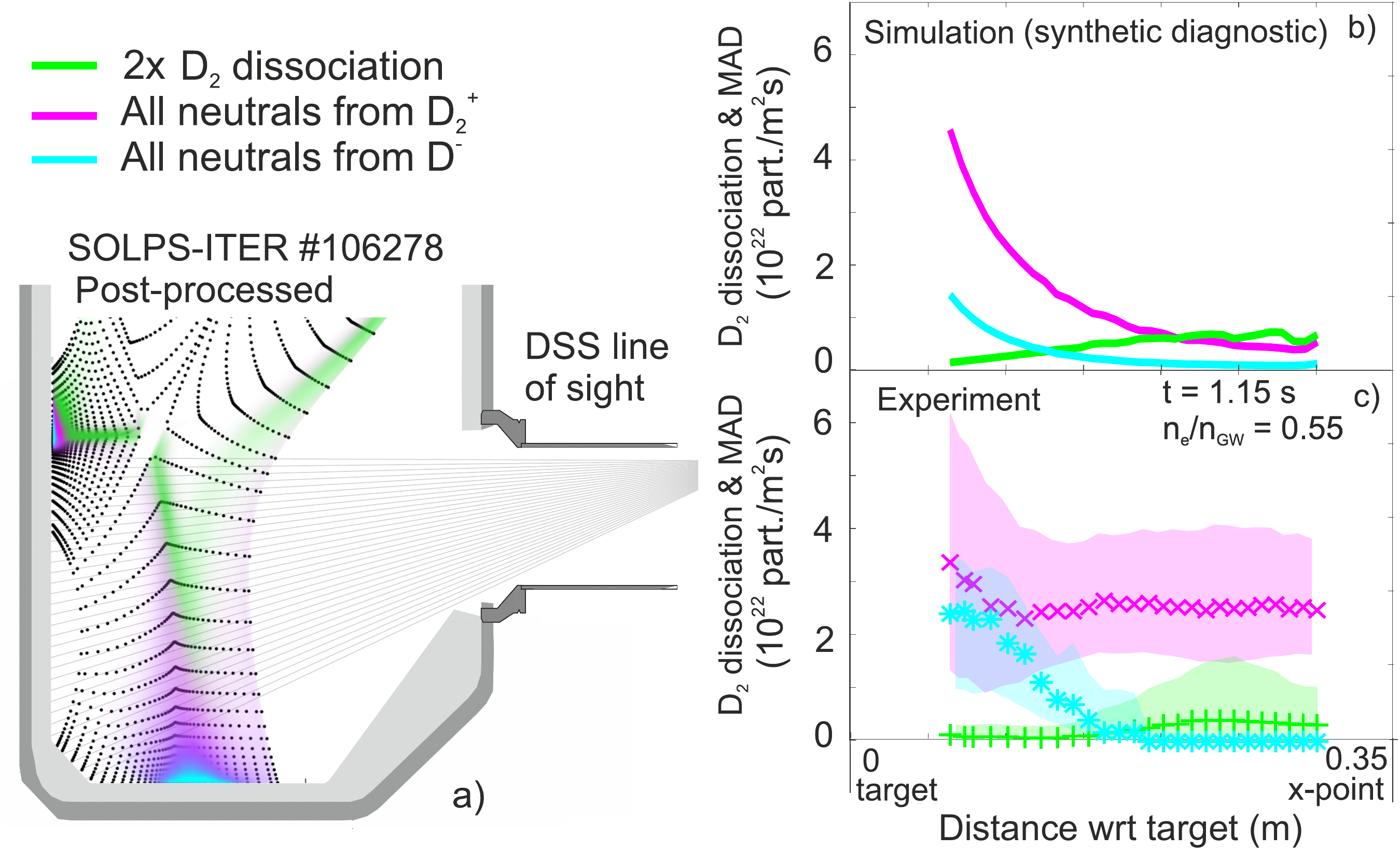}
    \caption{Overview of neutral atom sources due to '2x $D_2$ dissociation' (green) - which represents all neutrals generated from $D_2$ dissociation; 'All neutrals from $D_2^+$' (red) and 'All neutrals from $D^-$' (cyan). a) The 2D plot of dissociation neutral atom sources using SOLPS post-processing across the divertor cross-section - the opacity indicates the strength of the neutral atom sources; b) the 1D profile using synthetic chordal spectrometer measurements of the data in a); c) The results from analysis of experimental data using BaSPMI. }
    \label{fig:Dissociation}
\end{figure}

Using post-processing of the SOLPS-ITER simulation (sections \ref{ch:SOLPS_MAR}, \ref{ch:SOLPS_Dyna}), we compare the total quantity of neutrals arising from interactions with $D_2^+ \: \& \: D^-$ and $D_2$ (from $e^- + D_2 \rightarrow e^- + D + D$) in figure \ref{fig:Dissociation} a. This takes into account all neutral atoms created in the reaction chain - for instance $D_2^+$ MAR ($D^+ + D_2 \rightarrow D_2^+ + D$, followed by $D_2^+ + e^- \rightarrow D + D$) would result in 3 neutrals. We find that 98 \% of the neutrals arising from interactions with $D_2^+$ occur by MAD and MAR from $D^+ + D_2 \rightarrow D_2^+ + D$ followed by $e^- + D_2^+ \rightarrow D^+ + D$ and $e^- + D_2^+ \rightarrow D + D$, respectively. 

We find that the $D_2$ electron-impact dissociation region separates from the target during detachment  (figure \ref{fig:Dissociation} a). A comparison of figure \ref{fig:Dissociation} (b) with figure \ref{fig:Compa_Simulation_LoS}, shows the $D_2$ dissociation region movement follows a similar trend to that of the Fulcher band emission region. Based on post-processing the SOLPS-ITER results (figure \ref{fig:Dissociation} a,b), dissociation from $D_2^+$-related MAD remains close to the target and replenishes the loss of dissociation from $D_2$.  Plasma-molecule interactions with $D^-$ facilitate dissociation at yet lower temperatures. Integrating the dissociation rates figure \ref{fig:Dissociation} a for the outer divertor leg, we find that more than 85 \% of all neutrals from $D_2$ are generated through $D_2^+$ or $D^-$ in the outer divertor. Therefore, interactions with $D_2^+$, as well as possibly $D^-$, may also play a key role in determining the 2D neutral density profile below the ionisation region as lower temperature $D_2$ dissociation is enhanced.

The above sequence of dissociation events along the divertor leg as function of $T_e$ (e.g. electron-impact dissociation of $D_2$ followed by MAD of $D_2^+$ followed by MAD of $D^-$) can be explained by the molecular structure (see Fig. \ref{fig:MolRates}). Electron-impact dissociation of $D_2$ goes via excitation to an auto-dissociating triplet state ($b \: 3\Sigma u$), which at low vibrational levels is energetically similar to the upper level of the Fulcher band transition. Performing an analogous approach to figure \ref{fig:D2p_Rates} a,  one would expect dissociation to be strongest near 5 eV, while Fulcher emission would be strongest near 6 eV. $D_2$ ionisation ($e^- + D_2 \rightarrow 2 e^- + D_2^+$) would also start to occur at such temperatures (at 6 eV it has reached 70 \% of its peak level at 13 eV - figure \ref{fig:MolRates}). Therefore electron-impact dissociation, Fulcher band emission, $D_2$ ionisation are expected to occur roughly at similar locations in the plasma. Although all these molecular processes depend on the molecular density which strongly increases near the target, their reaction probabilities decrease so strongly at $T_e<5$ eV that their trend is effectively determined by the electron temperature rather than the molecular density (figure \ref{fig:Compa_Simulation_LoS}). A parallel to this is atomic ionisation, which occurs at similar temperatures: the ionisation region detaches from the target even though the neutral density is highest below the ionisation region. $D_2^+$ MAD is a dissociative attachment process and hence would occur at lower temperatures than electronic excitation dependent processes. $D^-$ MAD would occur at yet lower temperatures as $D^-$ is created through an electron attachment process, via $D_2^-$.


Using a similar method to that used to estimate MAR/MAI \cite{Verhaegh2021}, we estimate the amount of neutrals generated by reactions involving $D_2^+$ and $D^-$ (\ref{ch:MADCalc}, figure \ref{fig:Dissociation} c) and compare this against SOLPS-ITER modelling using synthetic diagnostics (figure \ref{fig:Dissociation} b). There is qualitative agreement between the MAD channels contribution to dissociation estimated experimentally and those from the synthetic diagnostic analysis of the post-processed SOLPS-ITER simulations for TCV, although the experimental estimates have a large uncertainty especially near the target. The experimentally inferred neutral source from MAD exceeds the simulated one closer towards the x-point ($>15$ cm from the target). However, here the MAD estimate from $D_2^+$ is likely overestimated as emission from electron-impact excitation may be misinterpreted as emission arising from interactions with $D_2^+$ (see \cite{Verhaegh2021a} for more explanation). Again we observe that the ratio between the neutrals generated from $D^-$ and those generated from $D_2^+$ is higher in the experimental case than in the SOLPS-ITER simulation. 

\subsection{Energy and power losses associated with MAR and MAD}
\label{ch:E_MAR}

Power losses (or gains) associated with ionisation and recombination depend both on radiative power losses, as well as whether potential energy has to be spent or is released by a reaction. Potential energy is gained by ionised particles after ionisation, which is then released upon surface recombination leading to additional wall power loading. However, during volumetric recombination, this potential energy is released back to the plasma. Whether recombination results in net power gains or losses then depends upon the balance between potential energy released back to the plasma and radiative losses during volumetric recombination through both MAR and EIR. To investigate this, we define the net effective energy loss ($>0$) or gain ($<0$) per recombination event as $E_{rec}$.


In \cite{Verhaegh2019} it was shown that electron-ion recombination on TCV resulted in $E_{rec}^{EIR} \approx 0$ eV, i.e. the radiative energy losses during electron-ion (radiative and three-body) recombination and the potential energy released roughly balance. However, at higher electron densities ($n_e>10^{20} m^{-3}$), as well as lower electron temperatures ($<1$ eV), relatively more recombination occurs through three-body rather than radiative recombination resulting in a reduced radiative loss (per recombination event) and thus net heating with $E_{rec}^{EIR} \geq -4$ eV \cite{Verhaegh2018}. 

\begin{figure}[H]
    \centering
    \includegraphics[width=\linewidth]{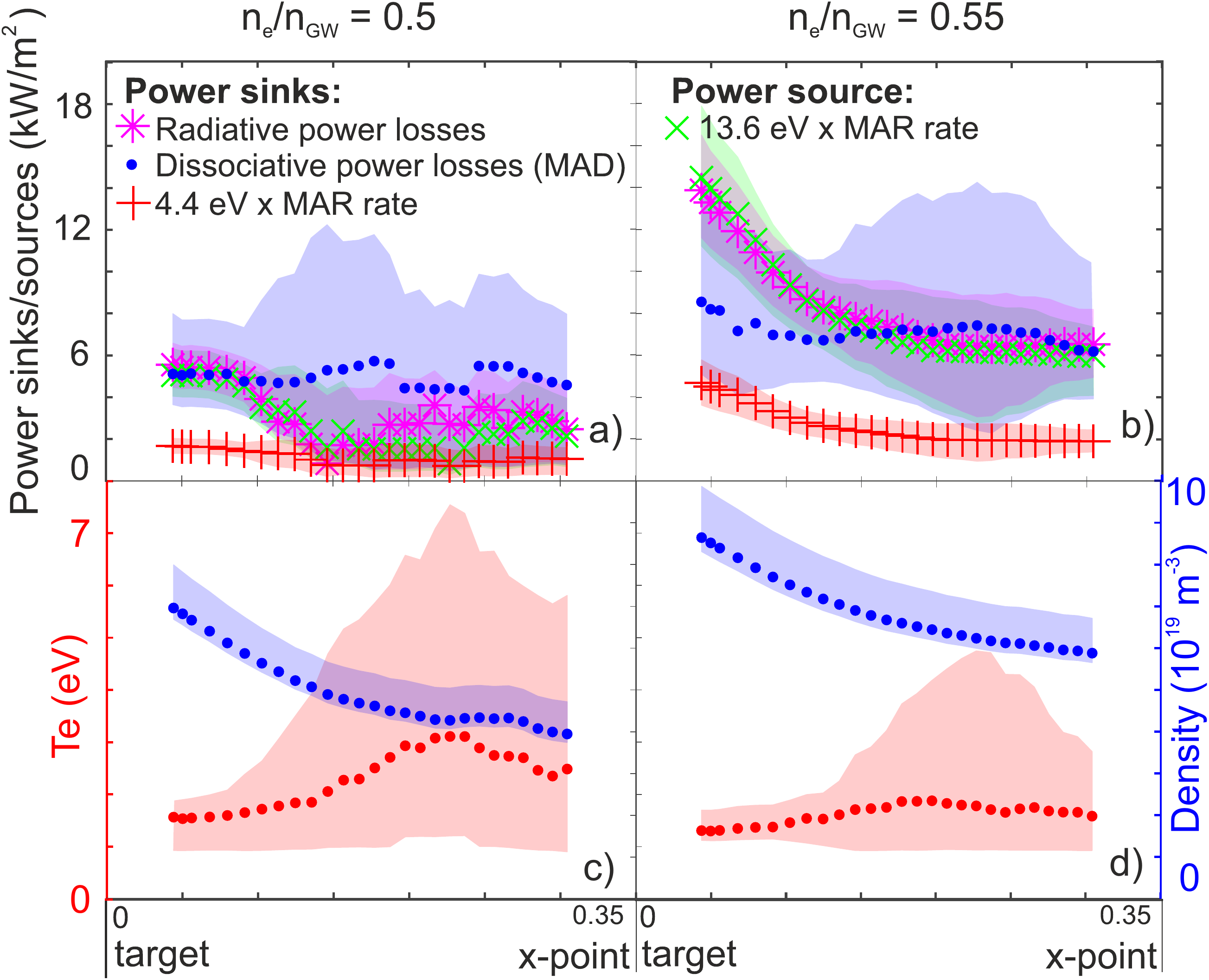}
    \caption{Profiles along the lines of sight of the inferred power sinks/sources attributed to $D_2^+$ and $D^-$ at two different density levels, corresponding to detachment onset (a) and detached (b): radiative power sink (magenta); power sink associated with MAD (2.2 eV per neutral) (blue); power source associated with the potential ion energy release during MAR (13.6 eV x MAR rate - green); power sink associated with dissociation \emph{during MAR} (4.4 eV x MAR rate - red). The shaded regions indicate the uncertainties in the inferences. c, d) estimated electron density (blue) and excitation temperature (red) profile using BaSPMI.}
    \label{fig:E_loss_Recomb_Ion}
\end{figure}

When we evaluate an equivalent $E_{rec}^{MAR}$, we obtain a similar result \emph{if we do not include dissociation}. Figure \ref{fig:E_loss_Recomb_Ion} shows profiles of radiative power losses associated with $D_2^+$ and $D^-$ at two different times in the discharge (magenta asterisk). Those radiative losses are shown to be approximately equal to the power released by the MAR ion sinks (green crosses), which is obtained by multiplying the MAR ion sinks with the potential energy released upon recombination. The radiative losses are mostly associated with MAR, but may also occur during MAD. If all radiative losses associated with $D_2^+$ and $D^-$ and the potential energy gain are considered, $E_{rec}^{MAR} \sim 0$ eV. 

\emph{However}, when we investigate the total reaction chains, these discussed MAR events also lead to the dissociation of molecules. Figure \ref{fig:E_loss_Recomb_Ion} also shows the power losses directly associated with dissociation during MAR, which would increase $E_{rec}^{MAR}$ by 4.4 eV. 

In addition to the dissociation directly associated with the MAR reaction, there is also molecular activated dissociation (MAD) (section \ref{ch:D2p_MAD}). The MAD power losses can be significant (figure \ref{fig:E_loss_Recomb_Ion}) and result in power losses in the cold plasma region below the $D_2$ dissociation region where MAD is particularly efficient (figure \ref{fig:Dissociation}). In a similar way to how radiative losses from electron-impact excitation collisions preceding ionisation are included in the ionisation energy losses, one could speculate that these MAD reactions occur before a MAR reaction occurs and, therefore, should be included in the MAR energy losses. This increases $E_{rec}^{MAR}$ to $10 \pm 5$ eV in the region with strongest MAR. In regions with lower MAR (e.g. higher up the divertor leg and/or earlier in the discharge) $E_{rec}^{MAR}$ can increase to $20 \pm 10$ eV as the ratio between dissociation and recombination is higher. We can thus conclude that energy losses due to dissociation during reactions with $D_2^+$ and $D^-$ can significantly elevate the MAR related energy losses, in contrast to EIR. These power losses are lower order estimates as collisions could transfer additional energy from the plasma to the molecules \cite{Myatra,Moulton2018}.

\subsection{The implications of our results}

\subsubsection{Potential differences between TCV and other devices}

The generality of these TCV results have to be further investigated as the impact of $D_2^+$ on detachment may be different on other devices. TCV was operated at relatively low power ($\sim 450$ kW, Ohmic) and low electron density ($\sim 10^{20} m^{-3}$), resulting in lower electron-ion recombination rates as well as relatively long mean free paths for atoms and molecules. As a result, we observe that ionisation does not occur in a thin region at the target but is spread throughout the divertor leg.  TCV features a nearly complete carbon wall. This means that the ratio between ions reflected as atoms vs recombination into molecules on the surface and emission from the surface is different than for a metal wall (higher reflection). The energy reflection coefficient of the recycled atoms will also be lower for a carbon wall than a metal wall. On the other hand, there are qualitative indications from JET (metal walls) measurements \cite{Lomanowski2020} of an enhanced $D_\alpha$ emission associated with MAR which is not accounted for in their simulations; that could mean that the JET divertor plasma is qualitatively consistent with our research findings.

Given the large mean free paths on TCV, the $D_2 (\nu)$ life-times could be sufficiently long for transport to play a role in the distribution of vibrational states, changing the likelihood of $D_2^+$ and $D^-$ generation \cite{Fantz2001}. A vibrational excited level $\nu$ will either be excited to $\nu+1$ or de-excite to $\nu-1$ \cite{Kotov2007}. Hence it can take a significant amount of reactions for highly vibrationally excited molecules to de-excite to lower vibrational levels, which makes the transport of vibrationally excited molecules more complex \cite{Fantz2001}. Such transport effects are likely more important than the transport of $D_2^+$, which is considered static in SOLPS-ITER simulations. This, combined with wall reflection effects, could change the vibrational distribution of the molecules leaving surfaces and in the plasma between experiments and simulations as well as between different devices. Such transport and reflection effects could possibly explain the inferred presence of $D^-$ related interactions, which requires a presence of highly vibrationally excited molecules. 

\subsubsection{The need for diagnosing and modelling the vibrational distribution}

It is clear from the discussions in this paper that more research is required to better understand and validate the effective molecular charge exchange rate used in plasma-edge codes, which is also sensitive to the vibrational distribution of the molecules. Molecular wall interactions, for instance, will also affect the vibrational distribution and are unaccounted for. A more detailed analysis including vibrational state resolved simulations that tracks the vibrational distribution together with plasma-wall interaction effects, could modify the simplified picture used in our simulations (section \ref{ch:SOLPS_MAR}). It may be possible to reduce the computational work-load required by bundling various vibrationally excited levels, which would require further investigation. 

Following the logic that the distribution of vibrational levels may need to be modelled more accurately, the improved match to experiment through the use of molecular charge exchange rates from \cite{Kukushkin2017} in post-processing SOLPS-ITER divertor plasma solutions may be ‘co-incidental’. The improvement may instead occur because higher vibrational levels are populated in the experiment than in SOLPS-ITER. 

All these aspects motivate investigating the transport of vibrational states in both experiments and simulations using diagnostics and synthetic diagnostics to compare against experiment. Since the Fulcher band emisison can be dim near the target in detached conditions \cite{Verhaegh2021a}, it should be investigated whether Fulcher emission is the best diagnostic of the vibrational distribution in strongly detached plasmas. 

\subsubsection{Implications of these TCV results for fusion reactors}

Our research indicated a lack of MAR from $D_2^+$ during $N_2$ seeded conditions and the reason for this was that the divertor target temperatures were higher than 3 eV. Since impurity seeding is of particular interest for reactors, the higher target temperatures during impurity seeding should be further investigated for a range of impurities. Given that MAR (1-3 eV) occurs in a higher temperature regime than EIR ($<1.5$ eV), MAR could potentially play an important role in divertor conditions were the target temperature is too high for EIR but sufficient for MAR. 

Previous work by Kukushkin \cite{Kukushkin2017} with SOLPS modelling of ITER plasmas indicates there can be a trade-off between MAR and EIR as the molecular rates are changed in the analysis. He found that with either the use of default, or modified, molecular charge exchange rates in SOLPS did not engender large differences in target heat flux and the ion saturation current. Instead, the electron-ion recombination rate reduced when MAR was included, cancelling the effect of increased MAR. Kukushkin attributes the reduction of electron-ion recombination to changes in the plasma cooling vs heating between three-body recombination, resulting in plasma heating ($E_{rec}<0$), and MAR. This additional heating was hypothesised to sustain more ionisation and could, therefore, result in higher electron densities and higher EIR rates. However, EIR does not have a net heating effect in lower electron density conditions, such as the ones shown in this study, which could impact those conclusions on current machines.

We find, through post-processing, that $D_2^+$ may facilitate dissociation at lower temperatures. This may be of particular importance for strongly detached scenarios, such as long-legged baffled divertor legs where the ionisation front can be held at a position far from the target \cite{Myatra,Moulton2018} - which may be reactor relevant \cite{Militello2021}. Self-consistent simulations are required to improve our understanding in a variety of conditions of how such molecular interactions play a role in reactors.

\subsubsection{The need for improvements in diagnostic techniques and direct measurements of $D_2^+$ and $D^-$}

One important caveat is that the results presented in this work rely on spectroscopic analysis using BaSPMI \cite{Verhaegh2021} and its various assumptions. Although the analysis has been thoroughly tested for avoiding the risk of 'overfitting' \cite{Verhaegh2021} and includes significant uncertainties on the atomic as well as molecular emission coefficients (15 and 30 \%), it provides inferences based on Balmer line intensities rather than direct measurements of the various molecular densities. It also relies on applying this data for hydrogen to a deuterium plasma as deuterium data is not available currently. More accurate and 2D inferences could be made using Bayesian analysis techniques using camera data \cite{Bowman2020,Perek2019submitted}.

Development of more direct diagnostics for the $D_2^+$ and $D^-$ content, using active spectroscopy, could provide more direct measurements of the atomic and molecular densities as well as the distribution of vibrational states \cite{Bacal2000,Tonegawa2003,Engeln2020,Vankan2004}. Such techniques, however, have generally not been applied to tokamaks as they require complicated setups. But they could be potentially applied to linear devices, where they could be supplemented with spectroscopic techniques \cite{Akkermans2020}.  More complete sets of the fundamental data for the isotopologues are starting to become available \cite{Scarlett2021}, which can be used by collisional-radiative models \cite{Wuenderlich2016,Wunderlich2020} to provide data for deuterium and tritium. Verifying these new (vibrationally-resolved) results \cite{Scarlett2021,Wunderlich2021} in dedicated experiments for $H, D$ and $T$, as well as using them in vibrationally resolved plasma-edge simulations, is crucial for determining what impact plasma-molecule interactions have on plasma-edge physics in fusion reactors.

\section{Conclusions}
\label{ch:conclusion} 

Analysis of experiment and modelling of TCV tokamak discharges presented herein show that plasma-molecule interactions can result in additional significant ion sources/sinks and power losses compared to just atomic processes and thus can have a strong role during divertor detachment. Most such interactions seem to involve $D_2^+$. However, evidence is also presented that $D^-$ may be present and play a significant role in power and particle balance, which implies the presence of highly vibrationally excited molecules. 

The ion target flux during plasma detachment is now observed (and modelled) to roll-over due to both a reduction in the divertor ion source linked to power limitation as well as an increase in molecular activated recombination (MAR). The latter only becomes significant for sufficiently low target temperatures ($<2.5$ eV), which was not reached during $N_2$ seeded induced detachment where, indeed, such effects were not observed. 

Detailed comparisons of experimental results against SOLPS-ITER simulations are presented that indicate plasma-molecule interactions involving $D_2^+$ are presently underestimated in SOLPS-ITER deuterium simulations when the default reaction set (vibrationally unresolved) is used. This may be related to the use of isotope rescalings of the effective molecular charge exchange rate in EIRENE (from hydrogen to deuterium) and could lead to strong underestimations of $D_2^+$ and associated power/particle losses. A lack of $D_2^+$ in simulations may also have implications for the spatial distribution of neutral particles, as well as further energy losses, in the divertor as interactions with $D_2^+$ (Molecular Activated Dissociation - MAD) can be an extremely efficient dissociation mechanism at low temperatures where the electron-impact dissociation of $D_2$ does not occur.

Further experimental investigations on other devices are required to investigate the occurrence of these processes on other devices, compare this against the TCV findings and assess more accurately the relevance of such interactions for fusion reactors. The observed mismatch of $D_2^+$ related interactions between experiment and simulations could have implications for fusion reactors with sufficiently deep detached operation where the ionisation region is held significantly above the target, as it could provide a power and particle sink (as well as a neutral atom source) at temperatures between 1.5 - 3.5 eV. 

\section{Acknowledgements}

Discussions with Detlev Reiter are kindly acknowledged and were very helpful. This work has received support from EPSRC Grants EP/T012250/1 and EP/N023846/1. It has been supported in part by the Swiss National Science Foundation and has been carried out within the framework of the EUROfusion Consortium and has received funding from the Euratom research and training programme 2014-2018 and 2019-2020 under grant agreement No 633053. The views and opinions expressed herein do not necessarily reflect those of the European Commission. The work by A. Kukushkin and A. Pshenov was supported by the Russian Science Foundation Grant No. 18-12-00329.

\section{References}

\bibliographystyle{elsarticle-num}
\bibliography{all_bib.bib}

\appendix

\section{Ion flow into outer divertor leg}
\label{ch:SOLPS_IUp}

The SOLPS-ITER simulations indicate an increase of the ion flow from upstream ionisation into the outer divertor leg upon detachment, as indicated by the mismatch between the red and green trend in figure \ref{fig:PowerPartBal} \cite{Fil2017,Fil2019submitted,Wensing2019}. This, in part, arises from neutrals escaping the divertor and ionising in the scrape-off layer upstream or in the core. With the agreement between the ion target flux and the ionisation source, previously it was argued that either the ion flow into the divertor was not significant or was cancelled by an additional ion sink. Our MAR estimates in the present study indicate a net ion flow into the divertor that increases at detachment onset and has, at the least, a similar magnitude to the MAR ion sink (equation \ref{eq:PartBal}). This is quantitatively consistent with $I_{up}$ from the SOLPS-ITER simulations. This implies that one loses high recycling conditions on TCV (without baffles) during detachment, which may also affect the balance between the inner/outer divertor in terms of neutral and ion transport \cite{Pshenov2017a}. 

\begin{figure}[H]
    \centering
    \includegraphics[width=\linewidth]{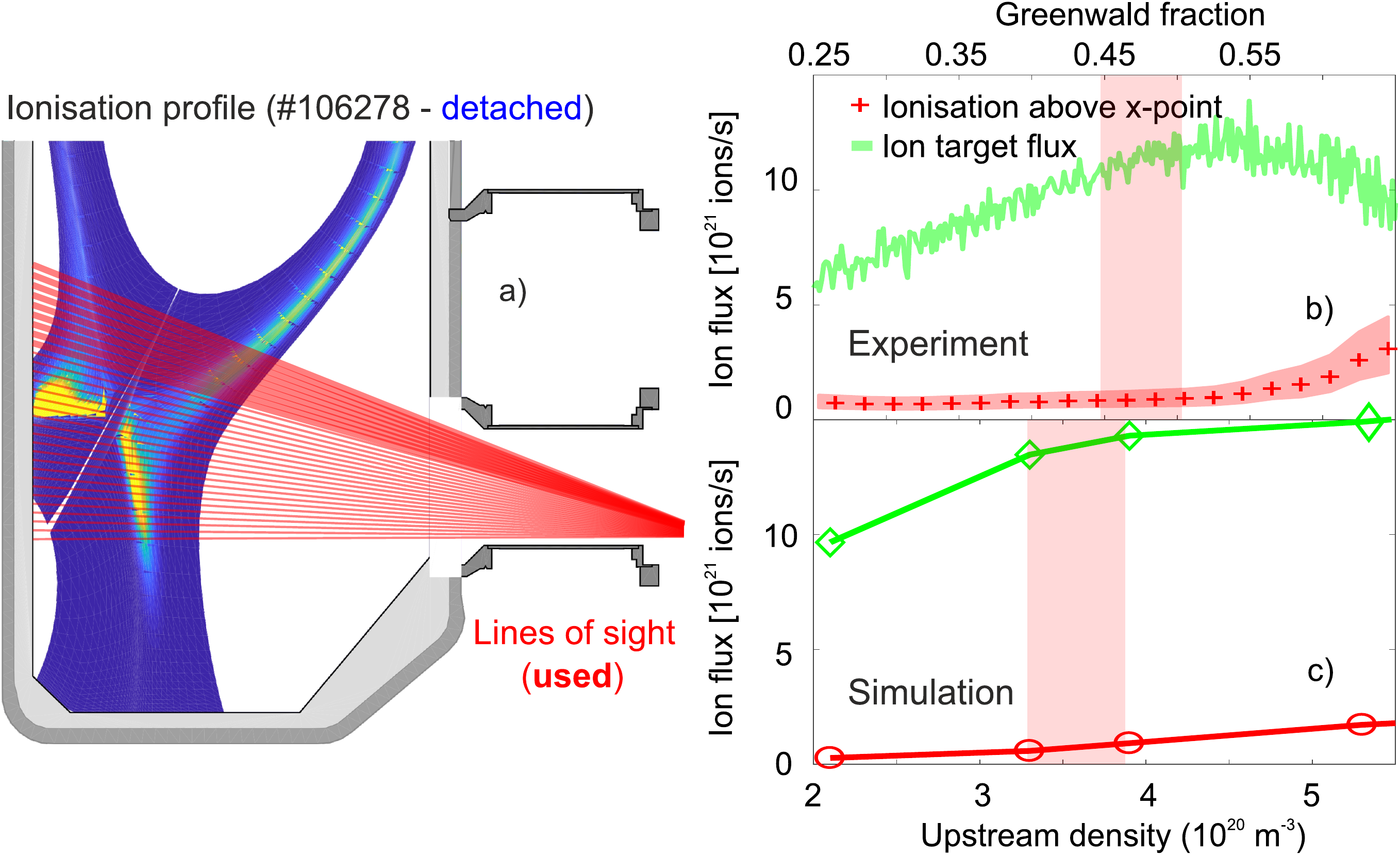}
    \caption{Investigation of ionisation above the x-point in the experiment and SOLPS-ITER. a) 2D ionisation profile from SOLPS-ITER at the detached phase (simulation \#106278 at an upstream density of $5.2 \times 10^{19} m^{-3}$ together with the lines of sight from the spectrometer, which have been tilted upwards to look above the x-point (the used lines of sight in the calculation are shaded)). b) Outer ion target flux and experimentally inferred ionisation source from the lines of sight (ions/s) upwards of the x-point as function of the Greenwald fraction (discharge \# 60403). c) Outer ion target flux (without post-processing) and ionisation source according to the lines of sight upwards of the x-point using a synthetic spectrometer for four different SOLPS-ITER simulations at four different levels of $D_2$ fuelling (and thus four different upstream densities) corresponding to attached (low and high density), detachment onset and detached conditions. For this calculation, only the ionisation at the open flux tubes have been considered (which is a factor 4-5 higher than the ionisation for the closed flux tubes intersected by the lines of sight). The vertical red shaded regions indicate the detachment onset.}
    \label{fig:UpstreamIoni}
\end{figure}

Figure \ref{fig:UpstreamIoni} a shows an increase of the ionisation upstream of the x-point during the detachment process in the simulations. In an attempt to monitor this upstream ionisation increase, the DSS spectroscopy views were rotated such that some viewing chords sampled above the x-point (see fig. \ref{fig:UpstreamIoni}a) and measurements of the $n=6,7$ Balmer lines were performed. Therefore, only the atomic aspects of the emission can be analysed under the assumption that plasma-molecule contributions can be neglected. Unfortunately, even though the repeat discharge, \# 60403, has qualitatively the same behaviour as \# 56567, the machine conditions were different (as the discharges were some months apart), resulting in a delayed detachment onset at a higher Greenwald fraction in the repeat discharge. As indicated by the displayed lines of sight, only a small portion of the additional upstream ionisation is probed, so a quantitative calculation of the total ion source outside of the lower divertor leg cannot be performed. 

The predicted measured upstream ionisation (figure \ref{fig:UpstreamIoni}b) is consistent with the simulated one (figure \ref{fig:UpstreamIoni} c). A caveat is that this measurement also records ionisation from near the inner target as well as from the core, as it is a line-integrated measurement. Ongoing studies using camera imaging data indicate an increase in upstream neutral density during detachment, qualitatively consistent with these findings.


More complex flow processes are also possible, where the ions are not just flowing from upstream towards the target, but also between the targets \cite{Pshenov2017a}. Spectroscopically, this cannot be probed as we can only detect a net effect of $\Gamma_u$ based on particle balance. Potentially coherence imaging measurements of ion flows \cite{Silburn2014} could provide more insight on this. In the performed simulations, the ion flow between the two strike points is negligible. Although the line-of-sight integral contributions to the ionisation for the closed flux tubes (figure \ref{fig:UpstreamIoni} a) are small, the ion flow from the core boundary increases by a factor two during detachment and can become non-negligible, although it remains smaller than the ion source in the upstream SOL.

\section{Balmer line emission associated with $D_2$ plasma chemistry and MAR/MAI with their respective temperature ranges}
\label{ch:MolTe}

\begin{figure}[H]
    \centering
    \includegraphics[width=\linewidth]{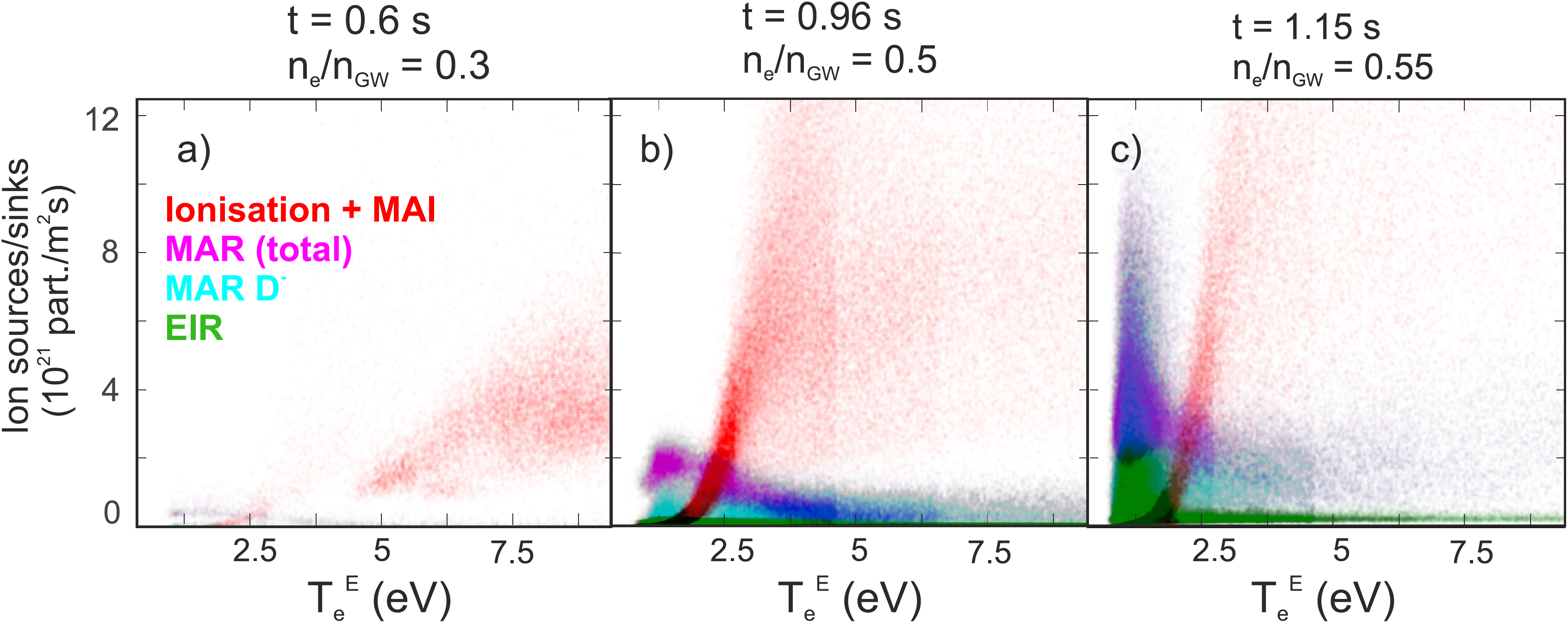}
    \caption{Overview of the relationship between the inferred $T_e^E$ and various ion sources and sinks for 3 time points in pulse \# 56567 corresponding to a density ramp to detachment for all lines of sight. Shown is the Monte-Carlo sampling output for the line integrated ionisation source (+ MAI) (red), total MAR (magenta), MAR ($D^-$) - cyan, EIR - green with transparency. The blue region is the overlap between the total MAR and MAR ($D^-$) regions (the colour magenta mixed with cyan leads to blue), and thus indicates a region where a significant portion of MAR arises from $D^-$. The transparency of the individual points (representing 20000 'dice rolls' per time frame/line of sight) is set according to their probability (determined by their weight according to temperature constraints - see \cite{Verhaegh2021}).}
    \label{fig:MC_samples_Te}
\end{figure}

Our MAR and temperature estimates (figures \ref{fig:PowerPartBal}, \ref{fig:N2_DenRamp_Compa})  are abstractions from the sample distributions obtained through Monte Carlo uncertainty propagation. However, there is a strong correlation within the samples between $T_e$ and parameters such as the ion source, EIR and MAR. To account for this while seeking MAR's $T_e$ regime, the sample outputs of $T_e^E$ (characteristic temperature of the electron-impact-excitation emission region along the line of sight obtained from BaSPMI analysis) are plotted against the various ion sources/sinks samples in figure \ref{fig:MC_samples_Te}  for the density ramp case (figure \ref{fig:PowerPartBal}). It is important to note that BaSPMI infers this from the different analysed brightness contributions: it does not use an assumed model for the $D_2^+/D_2$ ratio in this analysis. 

The results indicate MAR mostly occurs below 3 eV, whereas EIR starts to occur below 1.5 eV. One would expect, however, that MAR is reduced below 1 eV as there is insufficient energy to provide the vibrational excitation needed to promote $D_2^+$ creation (see \ref{ch:MolRates}). 

While a strong $T_e^E$ correlation with the ion source exists below 3 eV, this correlation mostly disappears above 4 eV. This explains the large uncertainty observed in the ion source in figures \ref{fig:PowerPartBal} and \ref{fig:PowerPartBalProfs}.

\section{'Effective' molecular rates}
\label{ch:MolRates}

\begin{figure}[H]
    \centering
    \includegraphics[width=\linewidth]{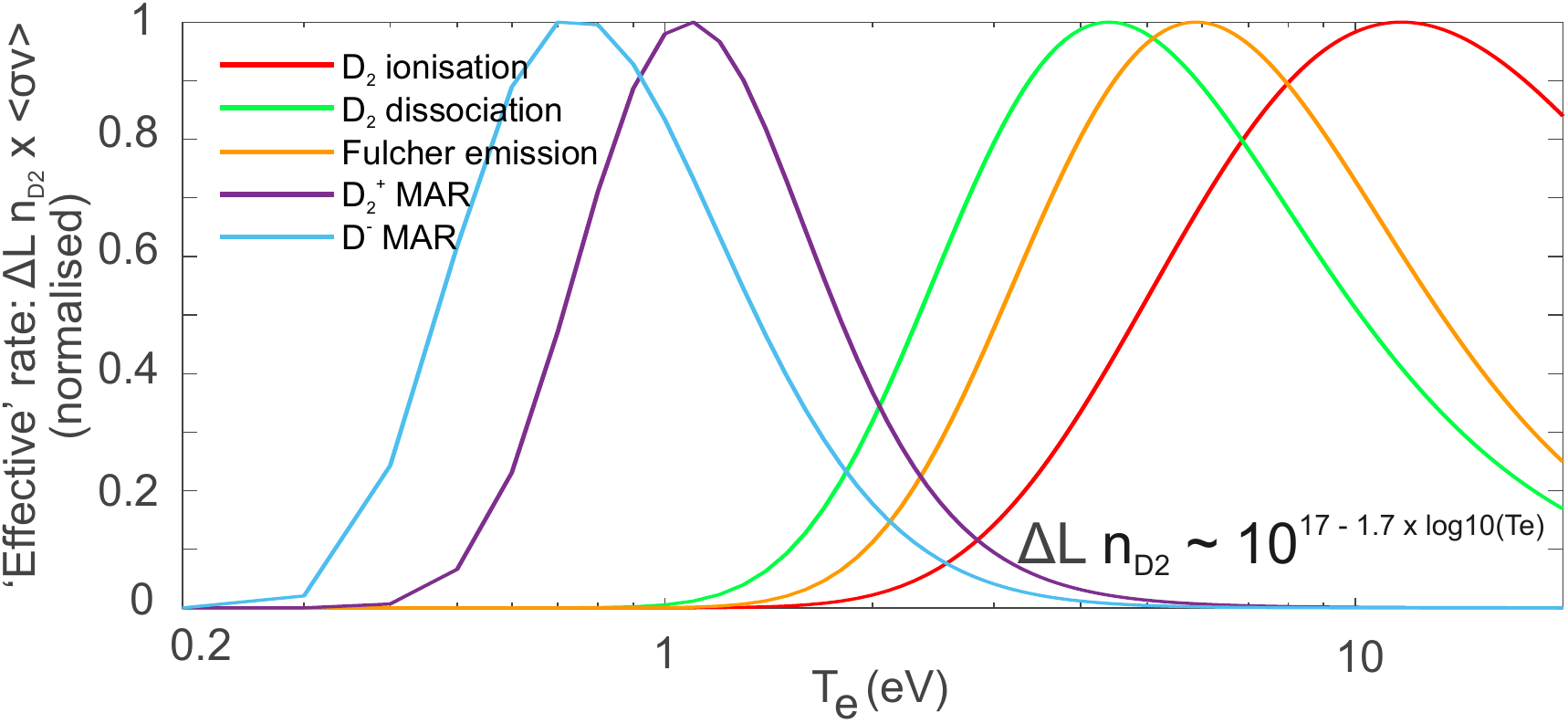}
    \caption{Normalised 'effective reaction rates' using AMJUEL \cite{AMJUEL} (e.g. $n_{D_{2}} \Delta L \times <\sigma v>$) as function of $T_e$, assuming $n_e = 5 \times 10^{19} \textrm{m}^{-3}$ and $n_{D_2} \propto 10^{17 - 1.7 \times \log{T_e}}$. For $D_2^+$ and $D^-$, their densities are determined using the $D_2$ density together with a modelled $D_2^+/D_2$ and $D^-/D_2$ ratio as function of $T_e$ using post-processing method nr. 3 (\ref{ch:PostProcessing}). Fulcher emission is obtained from AMJUEL rate H12 2.2.5fl \cite{AMJUEL}}
    \label{fig:MolRates}
\end{figure}

Analogously to the approach of figure \ref{fig:D2p_Rates}, a relation between the $D_2$ density from SOLPS-ITER \cite{Verhaegh2021a} has been multiplied with various $D_2$ interaction cross-sections in figure \ref{fig:MolRates}. This facilitates comparing the temperature regimes at which we suspect certain reaction mechanisms occur while taking into account the expected strong rise in the molecular density at low temperatures. This illustrates that $D_2$ ionisation ($e^- + D_2 \rightarrow 2 e^- + D_2^+$) is expected at the highest temperatures, followed by Fulcher band emission as well as electron-impact dissociation ($\sim 5$ eV or higher). Processes with $D_2^+$ are expected to be most dominant at $1-1.5$ eV, while processes with $D^-$ are suspected to be most dominant at $0.7-0.8$ eV.

\section{BaSPMI}
\label{ch:BaSPMI}

This work employs the BaSPMI analysis \cite{Verhaegh2019a,Verhaegh2021a} in order to infer information on both plasma-atom and plasma-molecule interactions from hydrogen Balmer line measurements. 

$D_\gamma$ and $D_\delta$ measurements are analysed initially assuming only 'atomic' (e.g. electron-impact excitation (EIE of $D$) and electron-ion recombination (EIR of $D^+$)) emission pathways of the hydrogenic line series allowing a prediction of the $D_\alpha$ and $D_\beta$ magnitudes. \emph{The difference between predicted and measured $D_\alpha, D_\beta$ is then ascribed to plasma-molecule interactions}. The strength of plasma-molecule interactions is then translated into molecule-derived contributions to $D_\gamma$ and $D_\delta$. This cyclic process is repeated until a self-consistent solution is found for all the Balmer line emission fractions due to electron-impact excitation (EIE), electron-ion recombination (EIR) and plasma-molecule related components ($D_2$, $D_2^+$, $D^-$) of $D_\alpha, D_\beta, D_\gamma$ and $D_\delta$. In this, the ratio between the plasma-molecule interaction contributions of $D_\beta$ and $D_\alpha$ is used to separate $D_2^+$ and $D^-$ related contributions. 

BaSPMI makes use of collisional-radiative model data from ADAS (atomic data) \cite{OMullane,Summers2006} and Yacora \cite{Wuenderlich2016,Wunderlich2020} (plasma-molecule interactions), as well as reaction rates from AMJUEL \cite{AMJUEL}. It is assumed that all this hydrogenic data can be directly applied to a deuterium plasma, which is further discussed in \cite{Verhaegh2021}. The technique uses Monte Carlo uncertainty processing, which is effective at catching non-unique solutions and accounting for these in the estimated uncertainty margins. Photon opacity effects have been assessed for this TCV discharge using SOLPS-ITER simulations and found to be negligible \cite{Verhaegh2021a}.

Previous research when developing BaSPMI \cite{Verhaegh2021,Verhaegh2021a} indicated that $D_3^+$ contributions to both the Balmer line emission as well as to particle losses is expected to be negligible. Post-processing (\ref{ch:PostProcessing}) using hydrogen rates indicated a contribution of $D_3^+$ to the Balmer line emission and particle losses of significantly below 1 \%.

Unless mentioned otherwise, the full BaSPMI approach is applied in this paper. However, there are various reductions (compared to the full analysis explained above) of the BaSPMI analysis possible.

\begin{enumerate}
\item BaSPMI can be employed without $D_\beta$ measurements, by omitting the step where the plasma-molecule interaction contribution to $D_\alpha$ (and $D_\beta$) is separated in $D_2^+$ and $D^-$ contributions. Instead, it could be assumed that all of this emission is attributed to $D_2^+$. This is employed as an alternative analysis in the text of section \ref{ch:detach_dyna}. However, the shown results in figure \ref{fig:PowerPartBalProfs} are from the full BaSPMI analysis.
\item BaSPMI can be employed without taking any plasma-molecule interactions into account (without $D_\alpha$ and $D_\beta$ measurements). Alternatively, the self-consistent treatment of the plasma-molecule interaction contributions to the medium-n (e.g. $D_\gamma, D_\delta$) Balmer lines can be omitted - which are then to be assumed due to plasma-atom interactions only. This can accurately reproduces the MAR rates (when $D_\alpha$ measurements are compared against the purely atomic extrapolation of $D_\gamma, D_\delta$), but may lead to overestimates of the ion source if MAR is significant \cite{Verhaegh2021a}. This is employed in section \ref{ch:N2_MAR}.
\end{enumerate}

\section{SOLPS-ITER post-processing}
\label{ch:PostProcessing}

In this work, existing SOLPS-ITER are post-processed to calculate alternative $D_2^+$ and also $D^-$ densities. This ignores the transport of $D_2^+$ and $D^-$ (which is also ignored in the default implementation of SOLPS-ITER). If transport is ignored, the $D_2^+/D_2$ ratio depends on the local balance between the rates of $D_2^+$ creation and the rates of $D_2^+$ destruction - assuming $n_e = n_{H^+}$ \cite{Verhaegh2019a}. The following creation/destruction rates are used: Creation rates: molecular charge exchange and $H_2$ ionisation $e^- + H_2 \rightarrow 2 e^- + H_2^+$ - AMJUEL H4 2.2.9 \cite{AMJUEL}. Destruction rates: $e^- + H_2^+ \rightarrow H + H$ - AMJUEL H4 2.2.11 \cite{AMJUEL}; $e^- + H_2^+ \rightarrow e^- + H + H^+$ - AMJUEL H4 2.2.12 \cite{AMJUEL} and $e^- + H_2^+ \rightarrow 2 e^- + H^+ + H^+$ - AMJUEL H4 2.2.14 \cite{AMJUEL}. For all electron-impact reactions the isotope dependence are neglected, in accordance with previous studies in \cite{Reiter2018}. Using this model, $D_2^+/D_2$ ratios are then obtained using four different rates for molecular charge exchange (e.g. $D_2 + D^+ \rightarrow D_2^+ + D$) - which are discussed in section \ref{ch:discuss_H2pHm_rate}, Fig. \ref{fig:D2p_Rates}:

\begin{enumerate}
    \item the default approach of SOLPS-ITER for deuterium 'Deuterium - default (EIRENE rescaled) rate' (figure \ref{fig:D2p_Rates}, blue), where $<\sigma v>_{eff} (T) \rightarrow <\sigma v>_{eff} (T/2)$ \cite{Kotov2007} (e.g. both the cross-sections as well as the distribution of the vibrational states undergo an isotope mass rescaling).
    \item the 'Deuterium - Kukushkin' rate from \cite{Kukushkin2017} (figure \ref{fig:D2p_Rates}, red), which does not take any chemical isotope differences into account for the molecular charge exchange, but only applies an isotope mass rescaling on the cross-sections. This is the default rate used for post-processing in this work. This leads to an order of magnitude difference compared to rate number 1 at 1-3 eV.
    \item the default hydrogen rate from AMJUEL  (figure \ref{fig:D2p_Rates},green). This leads to a 10-15 \% difference compared to rate number 2.
    \item the default hydrogen rate from AMJUEL multiplied with the deuterium to hydrogen ratio from \cite{Reiter2018} , which assumes the same vibrational distribution (e.g. Boltzmann) for deuterium and hydrogen but accounts for both isotope mass differences as well as chemical differences in the application of the cross-sections. This leads to a up to 3 \% difference compared to rate number 3.
\end{enumerate}

Those new $D_2^+/D_2$ ratios are calculated relative to the ratio obtained directly from SOLPS-ITER and this is used to rescale the $D_\alpha$ emissivity, MAR ($D_2^+$) and MAI ($D_2^+$) rate for each simulation grid cell, which are also used to modify the particle balance and through that the simulated ion target flux (sections \ref{ch:SOLPS_MAR}, \ref{ch:SOLPS_Dyna}). Note that the used models neglect interactions that occur in a mixed $HD$ plasma, which is considered in \cite{Reiter2018}.

As SOLPS-ITER does not include any interactions with $H^-$ or $D^-$ by default, $D^-$ is not included in the results in figure \ref{fig:Compa_Simulation_Integrated}. It is, however, included in sections \ref{ch:SOLPS_Dyna}, \ref{ch:D2p_MAD}. To investigate the role of $D^-$, the default hydrogen $H^-/H_2$ ratio from AMJUEL was rescaled to $D$ using coefficients from \cite{Reiter2018}, which reduces the $D^-$ density by $\sim 30 \%$. This, however, is strongly sensitive to the assumed model/distribution for the vibrational states $f_{\nu}$ as the creation cross-section for $D^-$ is strongly isotope dependent at low vibrational levels.


 
 \section{Estimating the neutral source from interactions with $D_2^+$ and $D^-$}
 \label{ch:MADCalc}
 

The MAD rates are estimated analogously to how the MAR rates are estimated by computing a 'MAD (or dissociation) per emitted $D_\alpha$ photon' ratio for $D_2$, $D_2^+$ as well as $D^-$. As SOLPS-ITER post-processing indicates that most (98 \%) neutrals generated through $D_2^+$ arise from MAR as well as molecular charge exchange followed by $e^- + D_2^+ \rightarrow e^- + D^+ + D$ (section \ref{ch:D2p_MAD}), only that particular chain is accounted for when determining the 'MAD per $D_\alpha$ photon' ratio for $D_2^+$. These 'MAD (or dissociation) per $D_\alpha$ photon' ratios are  multiplied by the respective contribution (e.g. $D_2$, $D_2^+$, $D^-$) to the $D_\alpha$ brightnesses to obtain an estimate of the quantity of neutrals created from MAD ($D_2^+$). The uncertainty margin for MAD is larger than for MAR due to the strong temperature sensitivity of the 'MAD per emitted $D_\alpha$ photon' ratio.













\end{document}